\documentclass[natbib]{svjour3}
\usepackage[dvips]{graphicx}
\newcommand{\eps}{\varepsilon}
\newcommand{\Dd}{\overline{D}}

\newcommand{\hte}{\widetilde{h}}
\newcommand{\hd}{\overline{h}}
\newcommand{\gd}{\overline{\vphantom{h}g}}
\newcommand{\Kt}{\widetilde{{\cal K}}}

\newcommand{\Kk}{{\cal K}}
\newcommand{\rb}{{\bf r}}
\newcommand{\eb}{{\bf e}}

\newcommand{\mod}{\mathop{\rm mod}\nolimits}
\newcommand{\vphi}{\varphi}
\newcommand{\vphid}{\overline{\vphantom{h}\varphi}}
\newcommand{\Pp}{P_\varphi}
\newcommand{\Pg}{P_g}
\newcommand{\Ph}{P_h}
\newcommand{\Phid}{\overline{\Phi}}
\newcommand{\taud}{\overline{\vphantom{h}\tau}}
\newcommand{\Td}{\overline{T}}

\newcommand{\xd}{\overline{\vphantom{h}x}}
\newcommand{\xid}{\overline{\vphantom{h}\xi}}

\newcommand{\yd}{\overline{\vphantom{h}y}}
\newcommand{\Wd}{\overline{W}}
\newcommand{\zetad}{\overline{\vphantom{h}\zeta}}
\newcommand{\tg}{\mathop{\rm tg}\nolimits}
\newcommand{\ctg}{\mathop{\rm ctg}\nolimits}
\newcommand\beq[1]{ \begin{equation}\label{#1} }
\newcommand{\eeq}{ \end{equation} }

\newcommand\beqa[1]{ \begin{eqnarray}\label{#1}}
\newcommand{\eeqa}{ \end{eqnarray} }
\newcommand{\beqano}{ \begin{eqnarray*} }
\newcommand{\eeqano}{ \end{eqnarray*} }
\journalname{Celestial Mechanics and Dynamical Astronomy}
 \title{Quasi-satellite orbits in the general context of dynamics in the 1:1 mean motion
 resonance. Perturbative treatment
\footnote{Results of this paper were partially presented as Paper DDA 101.04 at the 44th Annual Meeting of the
 Division on Dynamical Astronomy of the American Astronomical Society, 2013, Paraty, Brazil.}
}
\titlerunning{Quasi-satellite orbits. Perturbative treatment}
 \author{Vladislav V. Sidorenko \and
         Anatoly I. Neishtadt \and
         Anton V. Artemyev \and
         Lev M. Zelenyi}
 \institute{ V.V.Sidorenko
           \at Keldysh Institute of Applied Mathematics \\
               Russian Academy of Sciences, \\ \smallskip
               Miusskaya Sq., 4, 125047 Moscow, RUSSIA \\
               Moscow Institute of Physics and Technology \\
               Institutskiy S-Str., 9, 141700 Dolgoprudny, RUSSIA \\
               \email{vvsidorenko@list.ru}
            \and
             A.I.Neishtadt
           \at Loughborough University, \\
               LE11 3TU Loughborough, UK
               \and
             A.I.Neishtadt \and A.V.Artemyev \and L.M.Zelenyi
           \at Space Research Institute \\
               Russian Academy of Sciences, \\
               Profsoyuznaya Str., 84/32, 117997 Moscow, RUSSIA
}

 \authorrunning{V.V.Sidorenko et al.}
 \date{}
 
\begin{document}
 \bibliographystyle{plainnat}
 \maketitle
 \begin{abstract}
 Our investigation is motivated by the recent discovery of asteroids orbiting the Sun and simultaneously staying near one of
 the Solar System planets for a long time. This regime of motion is usually called the quasi-satellite regime, since even at the times of the closest approaches the distance between
 the asteroid and the planet is significantly larger than the region of space (the Hill's sphere) in which the planet
 can hold its satellites.
 We explore the properties of the quasi-satellite regimes in the context of the spatial restricted circular three-body problem ``Sun-planet-asteroid". Via double numerical averaging,
 we construct evolutionary equations which describe the long-term behaviour of the orbital elements of an asteroid. Special attention is paid to possible
 transitions between the motion in a quasi-satellite orbit and the one in another type of orbits available in the 1:1 resonance. A rough classification of the corresponding
 evolutionary paths is given for an asteroid's motion with a sufficiently small eccentricity and inclination.
 \end{abstract}

 \section{Introduction}

\begin{figure}[htb]
\vglue4cm
\hglue0.5cm
\includegraphics[width=0.75\textwidth,keepaspectratio]{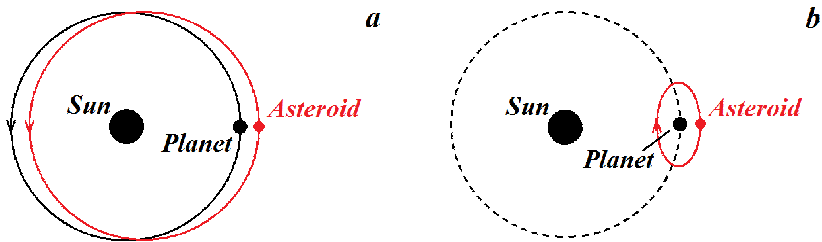}
\vglue-1.cm

\caption{The orbital motion of a quasi-satellite and its host planet. Panel $a$ is a Sun-centered
reference frame {\bf that preserves the orientation in the absolute space}. The quasi-satellite and the planet move around the Sun with the same orbital period in elliptic
and in circular orbits respectively. Panel
$b$ is a Sun-centered frame rotating with the mean orbital motion of the planet.}
\label{Pic0}
\end{figure}

 During recent decades, the properties of the so-called quasi-satellite orbits (QS-orbits) have been intensively explored. Within the scope of the three
 body problem ``Sun-planet-asteroid", the motion of an asteroid in a QS-orbit corresponds to the 1:1 mean motion resonance, with the resonant argument $\vphi=\lambda - \lambda'$ librating around $0$ ($\lambda$ and $\lambda'$ being the mean longitudes of the asteroid and the planet, respectively). The asteroid motion in QS-orbit is bounded to the planet's
 neighborhood of a size which can be small enough in comparison  with the value of semimajor axis $a'$ of the planet (Fig.~\ref{Pic0}). Nevertheless, the trajectory of the asteroid will never cross the Hill sphere of the planet, wherefore the asteroid cannot be considered as a satellite in the usual sense.

 To our knowledge, for the first time the existence of QS-orbits was discussed by \citet{Jackson1913}. {\bf Long enough the studies of this class of orbits were limited to the consideration of
 the periodic motions classified as ``f-family" by Str\"{o}mgren. As remarkable from the different points of view we can mention in this  context the investigations by~\citet{Broucke1968}, \citet{Henon1969} and \citet{Benest1977}.
 Carrying out his research on the NASA contract \citet{Broucke1968} actually anticipated the application of QS-orbits
 in astrodynamics. \citet{Henon1969} proved the (planar) stability of f-family in the Hill approximation and conjectured
 the existence of the natural retrograde "satellites" that are much farther from the host planet than the collinear libration
 points $L_1$ and $L_2$. \citet{Benest1977} established the conditions for three-dimensional stability of f-family periodic
 orbits in the frame CR3BP.

 At the end of 80th the opportunity to insert a spacecraft into QS-orbit around the Martian moon Phobos was
 thoroughly studied and finally realized in the former USSR \citep{Kogan1989, LidovandVashkovyak1993, LidovandVashkovyak1994}. Phobos was the goal
 of the last Soviet interplanetary mission \citep{SagdeevandZakharov1989}. Since the Hill sphere of Phobos is very
 close to its surface it is impossible to circumnavigate it in a Keplerian-type way. Inspired by the results of \citet{Henon1969}
 and \citet{Benest1976} A.Yu.Kogan (then a mission specialist at Lavochkin Aerospace company, USSR) proposed a QS-orbit
 as a solution. Although this mission was only partially successful one of the two launched spacecrafts attained
 stable QS-orbit (A.Yu.Kogan, private communication). The related activity was reviewed in \citep{Kogan1990}, where in
 particular the currently popular definition of QS-orbit was formulated probably for the first time: quasi-satellite orbits are the trajectories of restricted three-body problem which are (1) located far beyond the Hill's sphere surrounding the minor primary body and (2) much less distant from it than from the major primary. Later the application of QS-orbits in astrodynamics was considered by \citet{Tuchin2007}, \citet{GilandSchwartz2010} and many other specialists.}

 Since outside the Hill sphere the gravity field of the planet is weak enough, a QS-orbit can be treated as a slightly perturbed heliocentric Keplerian ellipse. It offers great opportunities for analytical consideration of the motion in QS-orbits. {\bf For the first time such a strategy was applied by \citet{MikkolaandInnanen1997}.
 A little bit later the perturbation approach was used to study the properties of OS-orbits in the papers by \citet{Namouni1999} and \citet{Namounietal1999}. These papers provided the greatest progress in understanding
 of the key dynamical structures responsible for the long-term evolution at 1:1 mean motion resonance. New types of orbital
 behavior were described (in particular, the so called compound orbits). The terminology (except for the minor modifications) and the concepts introduced in  \citep{Namouni1999} and \citep{Namounietal1999} became standard for further theoretical research of QS-orbits~\citep[][etc.]{Christou2000,Brasseretal2004,Mikkolaetal2006}\footnote{\bf Being evidently unaware about the studies of QS-orbits by specialists in astrodynamics, \citet{Namouni1999} and \citet{Namounietal1999} used for this type of the orbital motion the term "retrograde satellite orbit"  taken from the classical investigations of the periodic solutions in RC3BP. Looking through the literature it is
 easy to note that the community of celestial mechanicians is not uniform regarding what is more preferable.
 Since our activity was stimulated by the long standing discussions with specialists who coined the term "QS-orbit" in 80th
 (A.Yu.Kogan, M.L.Lidov and M.A.Vashkovyak), it predetermined our choice.}. Below we will be dealing also with many
 of these concepts.}

 In the last decade the interest to QS-orbits has increased due to the discovery of the actual quasi-satellites for
 Venus~\citep{Mikkolaetal2004}, Earth~\citep{Connorsetal2004, Wajer2009, Wajer2010}, Jupiter~\citep{KinoshitaandNakai2007} and
 Neptune~\citep{FuenteMarcosandFuenteMarcos2012}.

 The role of the quasi-satellite dynamics in the early Solar nebula was discussed by~\citet{Kortenkamp2005}.
 \citet{Giupponeetal2010} investigated the properties of the QS-motions in extrasolar planetary systems.

 An important phenomenon revealed by~\citet{Namouni1999} and \citet{Namounietal1999} is a possibility for an
 asteroid in the 1:1 mean motion resonance to change, from time to time, the qualitative character of its orbital motion. In particular, under a special choice of the initial conditions, transitions between the motion in
 a QS-orbit and that in a horseshoe orbit (HS-orbit) take place. More complicated scenaria are possible too~\citep{Brasseretal2004,Namouni1999,Namounietal1999}.

 To study  the secular evolution of the resonant motion with qualitative changes in the behavior of the argument $\vphi$ , one can apply an approach developed by ~\citet{Wisdom1985} in his investigation of the 3:1 mean motion resonance.
 This approach is, in fact, general enough and contained no restrictions on the type of the resonant orbit to model
 \citep[e.g.,][]{Yokoyama1996}. {\bf In essence it is based on the presence of the adiabatic invariant in the
 asteroid's dynamics at the resonance. Calculating level curves of the adiabatic invariant one can draw
 phase portraits characterizing the secular evolution of the motion. The consideration of the adiabatic invariance
 violation (due to the transitional phenomena) allows to understand the appearance of the chaos in asteroid's motion.}
 The first step in application  of Wisdom's method to the 1:1 resonance was carried out in~\citep{Nesvornyetal2002}.

 {\bf In the system's phase space the resonant phenomena are localized: they occur in narrow resonance
 regions~\citep{Arnoldetal2006}. It always interesting to compare the properties of the resonant motions with the properties of non-resonant motions when corresponding phase trajectories are close to the border of the resonance region.
 In the case of 1:1 mean motion resonance a lot of information for such a comparison can be found in the papers
 on the non-resonant motions of the asteroid and the planet body in the close
 orbit~\citep[e.g.,][]{LidovandZiglin1974,Gladman1993}.}

 The goal of our paper is twofold: (i) to establish
 the conditions at which the motion in QS-orbit is possible, and (ii)
 to explore {\bf when} this regime of orbital motion is perpetual and {\bf when it is temporary}.
 {\bf Wisdom's scheme of the mean-motion resonance analysis allows to notice what was not noticed in the previous studies on quasi-satellite dynamics.}

 Section 2 begins with the description of an averaging procedure used to determine the secular evolution in a mean motion resonance. The first averaging is
 carried out over the orbital motion, whereafter the phase variables are rescaled, and the problem is shaped into a form called a ``slow-fast" system (SF-system). This is a two degrees of freedom Hamiltonian system with the variables evolving at different rates: some variables are ``slow", while the other are ``fast". The second averaging is then performed over the ``fast" motions of the SF-system. This provides us the evolutionary equations describing the secular
 effects in the asteroid's motion.

 Section 3 is devoted to the analysis of these secular effects, for various regimes of motion. The transitions between different regimes of orbital motion
 (QS$\rightarrow$HS, HS$\rightarrow$QS, etc) are discussed. In Section 4, the consideration is restricted to motion in orbits with a small inclination and eccentricity. In this case, the asteroidal dynamics demonstrates some kind of scaling. Section 5 provides an example of the dynamics of an actual asteroid in a QS-orbit. In Section 6, the summary of the main results is presented. Details of the averaging procedures are elucidated in Appendices A and B.

\section{Double averaged motion equations for investigation of
dynamics at 1:1 mean-motion resonance}

\subsection{Averaging over orbital motions}

Through all stages of our analysis we shall use the motion equations written in the Hamiltonian form. The units are chosen so that the distance between the primaries (i.e., between the Sun and the planet) is unity, the sum of their masses is also unity, while the period of their rotation around the system's barycenter is $2\pi$. Since the mass of the planet $\mu$ is substantially smaller than the mass of the Sun, the quantity $\mu$ is a small parameter of the problem.

To start with, we introduce the Delaunay canonical variables~\citep{MurrayandDermott1999}
$$
L, G, H, l, g, h,
$$
where $l$ is the mean anomaly of the asteroid, the other variables being related to the
asteroid's osculating elements (the semimajor axis $a$, the eccentricity $e$, the inclination $i$, the argument of pericenter $\omega$, the longitude of the ascending node $\Omega$) by the formulae
$$
L=\sqrt{(1-\mu)a},\qquad G=L\sqrt{1-e^2},\qquad H=G\cos i,
$$
$$
g=\omega, \qquad h=\Omega.
$$

The canonical relations give the equations of motion of the asteroid:

$$\frac{dL}{dt} = - \frac{\partial \Kk}{\partial
l}\,,\quad \frac{dG}{dt} = - \frac{\partial \Kk}{\partial
g}\,,\quad \frac{dH}{dt} = - \frac{\partial \Kk}{\partial
h}\,,\quad
$$
$$
\frac{dl}{dt} =  \frac{\partial \Kk}{\partial L}\,,\quad
\frac{dg}{dt} =  \frac{\partial \Kk}{\partial G}\,,\quad
\frac{dh}{dt} =  \frac{\partial \Kk}{\partial H}\,,
$$
with the Hamiltonian $\Kk$ defined as:

\beq{ih}
 {\cal K} = -\frac{(1-\mu)^2}{2L^2} - \mu
R(L,G,H,l,g,h-\lambda'),
\eeq
$R$ being the disturbing
function. For the restricted circular three-body problem
the disturbing function admits the form
\beq{pf} R = \frac{1}{|\rb - \rb'|} -
(\rb,\rb') \eeq
with  $\rb=\rb(L,G,H,l,g,h)$ and $\rb'=\rb'(\lambda')$
being the position vectors of the asteroid and the planet relative
to the Sun. The brackets around the second term in (\ref{pf}) denote
the scalar product.

The Hamiltonian ${\cal K}$ is a function of the time $t$ through
its dependence upon the planet's mean longitude $\lambda'$ (in our units,
$\lambda'=t+\lambda'_0$). As it follows from the formula (\ref{ih}), it is
reasonable to replace the variable $h$ with the variable $\hte = h
- \lambda'$. The new system's Hamiltonian $\Kt$ will be
time-independent:
$$
\Kt = {\cal K}(L,G,H,l,g,\hte) - H.
$$

Our next step is to perform the canonical transformation

\beq{ct1} (L,G,H,l,g,\hte)\longmapsto (\Pp, \Pg, \Ph, \vphi, \gd,
\hd) \eeq
defined by the generating function
$$
S = \Pp l + [\Pg+\Pp-1]g + [\Ph+ \Pp-1]\hte.
$$

The relations between new and old variables are:
$$
\begin{array}{ll}
L = \frac{\displaystyle \partial S}{\rule{0pt}{3.5mm}
\displaystyle
\partial l} = \Pp, & \vphi = \frac{\displaystyle\partial
S}{\rule{0pt}{3.5mm}\displaystyle \partial \Pp}
= l + \hte + g, \\
\rule{0pt}{8mm} G = \frac{\displaystyle \partial
S}{\rule{0pt}{3.5mm} \displaystyle
\partial g} = \Pg + \Pp - 1, &
\gd = \frac{\displaystyle \partial S}{\rule{0pt}{3.5mm}
\displaystyle\partial \Pg} = g, \\
\rule{0pt}{8mm} H = \frac{\displaystyle \partial
S}{\rule{0pt}{4mm} \displaystyle \partial \hte} = \Ph + \Pp - 1, &
\hd = \frac{\displaystyle \partial
S}{\rule{0pt}{4mm}\displaystyle
\partial \Ph} = \hte.
\end{array}
$$
 The purpose of transformation (\ref{ct1}) is to introduce the resonant phase $\vphi$ into the consideration. It is straightforward to verify that
$$
\vphi = l + (h - \lambda') + g = \lambda - \lambda',
$$
where $\lambda=l+h+g$ is the asteroid's mean longitude.

Once the transformation (\ref{ct1}) will be accomplished, the Hamiltonian assumes the form of
$$
\Kt = - \frac{(1-\mu)^2}{2\Pp^2} - \Ph - \Pp - \mu R~,
$$
where an insignificant constant term has been dropped.

Let ${\cal R}$ be a region in the system's phase space, defined by the condition

\beq{rr}
|\Pp - 1| \mbox{\raisebox{2pt}{$\mathop{<}\limits_{\displaystyle \sim}$}}\mu^{1/2}.
\eeq
\noindent We shall call it the resonant region, since the inequality (\ref{rr}) is equivalent to the
inequality
\begin{center}
$|n - n'|$\raisebox{2pt}{$\mathop{<}\limits_{\displaystyle \sim}$}$\mu^{1/2}$,
\end{center}
\noindent where $n$ and $n'$ are the mean motions of the asteroid and of the planet respectively \footnote
{\bf The given definition of the resonance region is a standard one for
the investigations of the resonant phenomena in multifrequency Hamiltonian system obtained from integrable one by a small perturbation of order $\mu$ \citep{Arnoldetal2006}. The asteroid entering into the Hill sphere results in the violation
of the last condition (if the integrable Hamiltonian corresponds to asteroid's motion in Keplerian orbit around Sun). For this reason we will consider only those trajectories which are definitely away from the planet's Hill sphere.
The perturbation theory should be developed in quite another way if one would like to take into account the dynamical effects
close or inside the Hill's sphere \citep[e.g.][]{RobutelandPousse2013}.}.

In the resonant region, the phase variables evolve at different rates. The variables $\Pp, \Pg, \Ph, \gd$ are the ``slow" ones:
$$
\frac{d\Pp}{dt},\;\frac{d\Pg}{dt},\;\frac{d\Ph}{dt},\;\frac{d\gd}{dt} \sim \mu~,
$$
 while the variable $\vphi$ is the ``semi-fast":
$$
\frac{d\vphi}{dt} \sim \mu^{1/2}~,
$$
 with $\hd$ being the only ``fast" variable in ${\cal R}$:
$$
\frac{d\hd}{dt} \sim 1~.
$$

As usual, the investigation of the secular effects in the motion of the asteroid
begins with the averaging over the fast variable. The thus-averaged Hamiltonian
$$
\Kt_{avr} = \frac{1}{2\pi}\int^{2\pi}_0 \Kt\, d\hd
$$
corresponds to the Hamiltonian system with two degrees of
freedom, and it depends upon $\Ph$ as a parameter.

Instead of the variables $\Pg, \gd$, it is now convenient to introduce the variables

\beq{xyd} x=\sqrt{2(1 - \Pg)}\cos \gd~,~\;
y=-\sqrt{2(1-\Pg)}\sin\gd~. \eeq

\noindent They are defined on the disk

$$
{\cal D}(P_h)=\{x^2 + y^2 \le 2(1 - P_h)\}.
$$
\noindent With an accuracy of $O(\mu^{1/2})$, the relations between $x, y, \Ph$ and the osculating eccentricity and inclination take
the form of

\beq{oe} e^2 = \frac{1}{4}(x^2 + y^2)[4 - (x^2+y^2)]~, \eeq
$$
\cos i = \frac{2\Ph}{2 - (x^2+y^2)}~.
$$
\noindent From Eq. (\ref{oe}), it follows that the centre of the disk ${\cal D}(P_h)$
corresponds to the motion in a circular orbit.

A similar relation for the argument of the pericenter can be written as
\beq{om}
\omega=\left\{
\begin{array}{cc}
2\pi - \arccos \frac{\displaystyle x}{\displaystyle \sqrt{x^2 +
y^2}}, & y \geq 0
\\
\rule{0pt}{8mm} \arccos \frac{\displaystyle x}{\displaystyle
\sqrt{x^2 + y^2}}, & y < 0
\end{array} \right.
(x^2+y^2 \ne 0).
\eeq

For a given value of $\Ph$, the eccentricity $e$ and the inclination $i$ of the asteroid do not exceed, at any moment of time, the values
 \beq{emax}
 e_{max} = \sqrt{1 - \Ph^2}+O(\mu^{1/2})
 \eeq
and
 \beq{imax}
 i_{max}=\arccos {P_h}+O(\mu^{1/2})~,
 \eeq
respectively.

 It would be worth dwelling upon the special case of $\Ph \approx 1$. As follows from relations (\ref{emax})-(\ref{imax}),
 this condition implies
$$
e_{max}\ll 1,\quad i_{max}\ll 1~.
$$
Therefore this  is the situation where the asteroid moves in the orbit of both a small eccentricity and a small inclination (Section 4).
 To simplify application of the perturbation technique to the analysis of the long-term evolution of motion, it will be instrumental to introduce
 the auxiliary quantity
$$
\sigma=\sqrt{1-\Ph^2}~.
$$
 Under the condition $\Ph \approx 1$, we clearly have
\beq{sigma}
\sigma \approx \sqrt{e^2+i^2} \ll 1.
\eeq
As it follows from (\ref{sigma}), the quantity $\sigma$
characterizes in a certain sense how close the orbit of the asteroid is to the orbit of the planet. At $\sigma \rightarrow 0$
the orbit of the asteroid becomes more and more close to the orbit of the planet.

\subsection{The ``slow-fast" system}

 Our next step is standard for the analysis of resonant phenomena in the multi-frequency
 systems~\citep{Arnoldetal2006}. Following the prescription of this theory, we undertake the scale transformation
$$
\tau = \eps t,\quad \Phi = (1 - \Pp)/\eps\approx (1-a)/2\eps~,
$$
where $\eps = \mu^{1/2}\,$. Without loss of accuracy, it is possible to rewrite the averaged equations of motion in ${\cal
R}$ as follows:

\beq{SF} \frac{d\vphi}{d\tau}= 3 \Phi,\quad \frac{d\Phi}{d\tau} = - \frac{\partial W}{\partial
\vphi}~, \eeq
$$
\frac{dx}{d\tau} = \eps \frac{\partial W}{\partial y},\quad
\frac{dy}{d\tau}= - \eps \frac{\partial W}{\partial x}~.
$$
Here
$$
W(\vphi,x,y,\Ph)=
$$
$$
=\frac{1}{2\pi}\int^{2\pi}_0 R[1,\Pg(x,y),\Ph,\vphi - \gd(x,y) -
\hd,\gd(x,y),\hd]\,d\hd.
$$
The expressions for $\Pg(x,y)$ and $\gd(x,y)$ can be obtained easily from Eq. (\ref{xyd}). {\bf Following \citep{Schubart1964} we apply the numerical integration to compute the averaged disturbing function $W(\vphi,x,y,\Ph)$.\footnote{Some authors
use for $W(\vphi,x,y,\Ph)$ the term ``ponderomotive potential" following~\citep{Namounietal1999}.}}

 The dynamical system (\ref{SF}) will be called below the ``slow-fast" system (or SF-system). Thus we would like to emphasise the existence of a
 timescale separation: the equations describing the behaviour of the variables $\vphi, \Phi$ make a ``fast" subsystem ($\frac{d\vphi}{d\tau},\frac{d\Phi}{d\tau}\sim 1$ in general), while the ``slow" subsystem consists of the equations for the variables $x,y$
($\frac{dx}{d\tau},\frac{dy}{d\tau}\sim \eps$).

{\it Remark.} Previously, the variable $\vphi$ and its conjugate momentum were classified as ``semi-fast" (Sec. 2.1). Averaging allows us to ``forget" about the processes on the orbital motion time scale. So in the subsequent analysis of the evolutionary equations (\ref{SF}) we shall regard $\vphi$ as a ``fast" variable without risk of a confusion.

The differential form
$$
\Psi = \eps^{-1} dy\wedge dx + d\Phi\wedge d\vphi
$$
defines a symplectic structure in the phase space of the SF-system. The Hamiltonian of this system is

$$
\Xi = \frac{3 \Phi^2}{2} + W(\vphi,x,y,\Ph).
$$

Finally, it is worth mentioning that differentiation of the averaged disturbing function $W(\vphi,x,y,\Ph)$ with respect to $\Ph$ results in the following equation describing the evolution of the longitude of the ascending node:
$$
\frac{d\Omega}{d\tau}=-\eps\frac{\partial W}{\partial \Ph}~.
$$

\subsection{Properties of the fast subsystem}

{\bf Some of the results discussed in this section are not new - they were obtained
in~\citep{Namounietal1999}. We rederive them here not only for the purposes of
self-containedness: to apply Wisdom's approach in the following sections we need
to know more about the properties of the fast subsystem than we were able to learn
from the preceding publications.}

At $\eps=0$, the fast dynamics is governed by the one-degree-of-freedom
Hamiltonian system
\beq{FS} \frac{d\vphi}{d\tau}= 3 \Phi, \quad
\frac{d\Phi}{d\tau} = - \frac{\partial
W}{\partial \vphi}~, \eeq
which depends on $x,y,\Ph$ as parameters. Let
\beq{sol}
\vphi(\tau,x,y,\Ph,\xi)\,,\;\Phi(\tau,x,y,\Ph,\xi)
\eeq
denote a solution of equations
(\ref{FS}), laying on the level set $\Xi = \xi$:
$$
\Xi(\vphi(\tau,x,y,\Ph,\xi),\Phi(\tau,x,y,\Ph,\xi),x,y,\Ph)=\xi.
$$
In general, the angle $\vphi$ oscillates or rotates:
$$
\vphi(\tau+T,x,y,\Ph,\xi) = \vphi(\tau,x,y,\Ph,\xi)\mod 2\pi~,
$$
$T(x,y,\Ph,\xi)$ being the fast motion period.

The solution~(\ref{sol}) is rotational in the case of
\beq{dp}
\max_\vphi W(\vphi,x,y,\Ph)< \xi
\eeq
The rotation of the resonant phase $\vphi$ corresponds to the motion of the
asteroid in the non-resonant  orbit. {\bf Following \citep{Namounietal1999}} we shall refer to it as the passing orbit or, briefly, the P-orbit. The set of points
$(x,y)$ satisfying inequality~(\ref{dp}) for given $\xi,\Ph$
will be denoted with $D_P(\xi,\Ph)$.

 Interpretation of oscillatory solutions requires caution since on the same level set
 $\Xi=\xi$ different types of such solutions may exist. Emergence of such variety depends upon the
 number and location of the local maxima and minima of the function $W(\vphi,x,y,\Ph)$, for
 given values of the ``parameters" $x,y,\Ph$.

\begin{figure}[htb]
\vglue-6.5cm \hglue3.5cm
\includegraphics[width=0.75\textwidth,keepaspectratio]{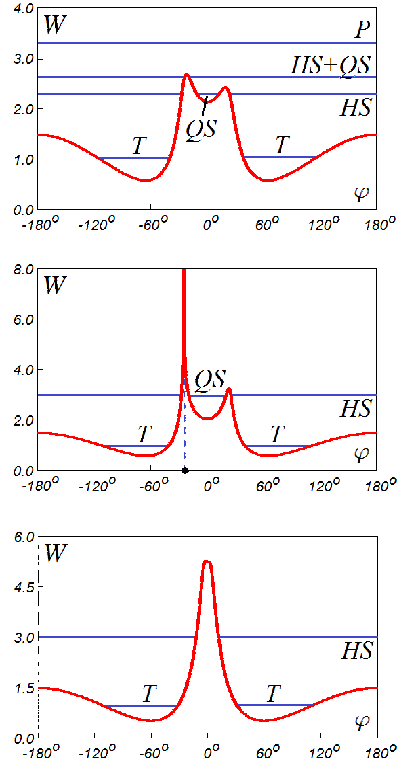}
\vglue1cm

\caption{Behaviour of the averaged disturbing function $W$ with fixed $x,y,\Ph$.
The following values are taken: $\Ph=0.96825$($\sigma=0.25$) in all cases,
$x=0.18509,y=-0.10686$ ($e=0.21250,\omega=30^\circ,i=7.74554^\circ$) in the upper panel,
$x=0.04542,y=-0.20884$ ($e=0.21250,\omega=77.73110^\circ,i=7.74554^\circ$) in the middle panel,
$x=0.05415,y=-0.03127$ ($e=0.06250,\omega=30^\circ,i=14.03624^\circ$) in the lower panel}
\label{Pic1}
\end{figure}

 To illustrate the situation, we present, in Fig.~\ref{Pic1}, several examples
 demonstrating various types of behaviour of the function $W$. {\bf Similar to~\citep[][Fig.2]{Garfinkel1977} and
 \citep[][Fig.1]{Namounietal1999}} the abbreviations near the
 level lines characterize the secular evolution of the corresponding motion of the
 asteroid on the intermediate time scale (i.e., on the interval of order $1/\mu^{1/2}$ in
 the initial units of time): QS - quasi-satellite orbit, HS - horseshoe orbit, P - passing orbit,
 T - tadpole orbit, QS+HS - a certain compound orbit {\bf (the existence of a variety of compound
 orbits at 1:1 mean motion resonance was revealed by~\citet{Namounietal1999}, more details can be found in
 \citep{Namouni1999, Christou2000}; first example of QS+HS orbit appeared  in~\citep{Wiegertetal1998})}. All of the above mentioned orbits (except for the P-orbit only) are described by oscillatory solutions.

\begin{figure}[htb]
\vglue4cm
\hglue0.5cm
\includegraphics[width=0.75\textwidth,keepaspectratio]{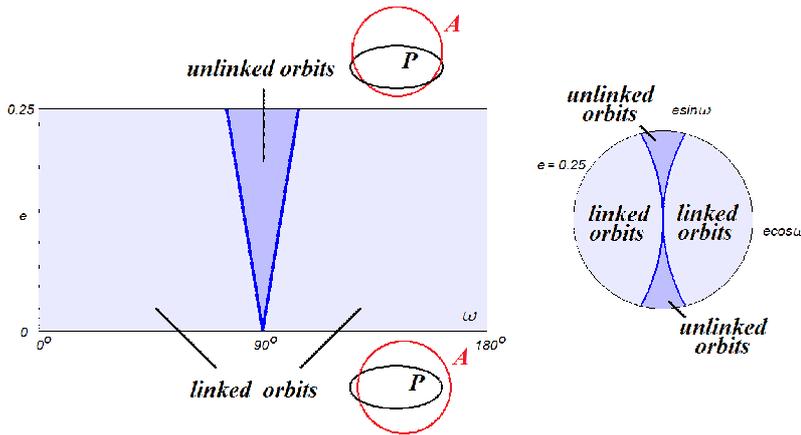}
\vglue-1.cm

\caption{ The partition of the set of asteroid's orbits ($n=n'$) provided by
the crossing condition (\ref{occ}): linked orbits - one node inside the planet's
orbit, the other is outside; unlinked orbits - both nodes inside the planet's
orbit. The blue line is a crossing curve defined by (\ref{occ}).
The diagram is drown in two ways. In what follows we find it is more
clear to present the phase portraits in the plane $(e,\omega)$. Circular diagrams are
convenient when we need to show regions satisfying a certain condition.}
\label{Pic_cc}
\end{figure}

 The middle panel in Fig.~\ref{Pic1} corresponds
 to the asteroid moving in the orbit which crosses the planetary orbit (the function $W$ thus being
 unlimited). {\bf In the limit $\mu=0$ the orbits crossing at the exact mean motion resonance ($n=n'$)  takes
 place when asteroid's osculating elements satisfy the condition}
\beq{occ}
\cos \omega = e.
\eeq
{\bf On the plane $(\omega,e)$ the crossing condition~(\ref{occ}) defines a curve, separating asteroid's
orbits linked and unlinked with the planet's orbit (Fig.~\ref{Pic_cc}). The motion in the crossing
orbits without a collision is possible due to the resonance.}

\begin{figure}[htb]
\vglue2cm
\hglue2.5cm
\includegraphics[width=0.5\textwidth,keepaspectratio]{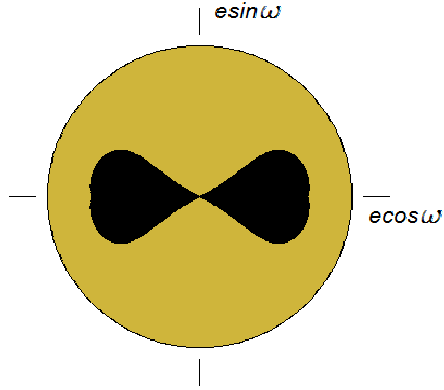}

\caption{The diagram demonstrates at what values of the slow variables the QS-motion is
possible for a given value of $\Ph=0.96825$($\sigma=e_{max}=0.25$). The set ${\cal B}_{QS}(\Ph)$
is painted in dark yellow.}
\label{Pic2}
\end{figure}

\begin{figure}[htb]
\vglue4.cm \hglue0.5cm
\includegraphics[width=0.75\textwidth,keepaspectratio]{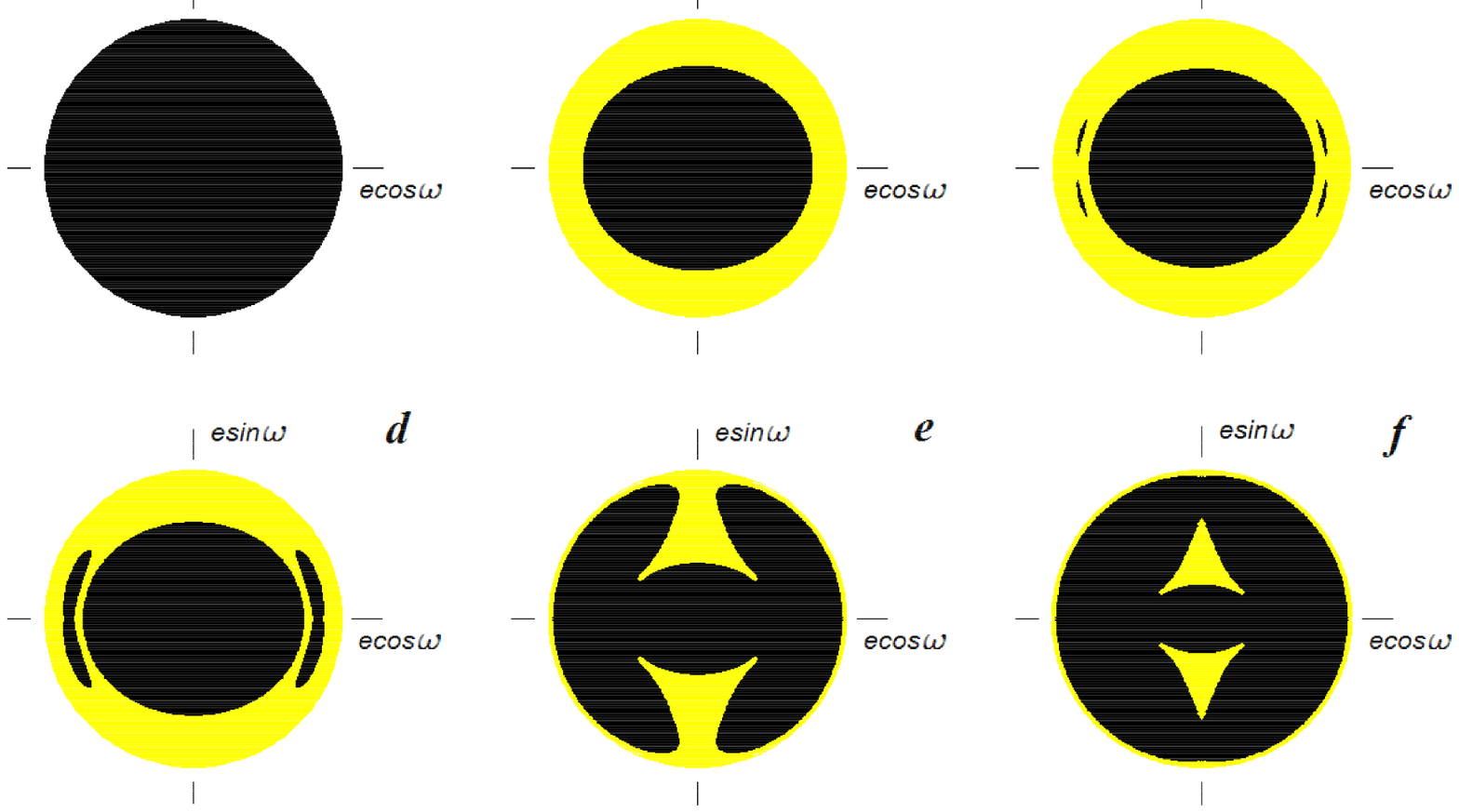}
\vglue-1.5cm 

\caption{A possible structure of the set $D_{QS}(\Ph,\xi)$ for
$\Ph=0.96825$} \label{diagram1} ($\xi_{min}=1.7735$, $\xi_h=2.3849$, $\xi_b=2.5470$,
$\xi_s=4.0606$): a - $\xi < \xi_{min}$ ($D_{QS}(\xi,\Ph)=\oslash$),
b - $\xi=2.35 \in (\xi_{min},\xi_h)$,
c - $\xi=2.40 \in (\xi_h,\xi_b)$, d - $\xi=2.45 \in (\xi_h,\xi_b)$,
e - $\xi=3.50 \in (\xi_b,\xi_s)$, f - $\xi=4.50 > \xi_s$
\end{figure}

 The lower panel in Fig.~\ref{Pic1} demonstrates that at some values of $x,y,\Ph$ motion in a
 QS-orbit is impossible. The relation between regions where such regimes are possible and the
 regions where such regimes are impossible is illustrated by Figs~\ref{Pic2},\ref{diagram1}. Specifically,
 Fig.~\ref{Pic2} provides the examples of the set
 ${\cal B}_{QS}(\Ph)\subset {\cal D}(\Ph)$ consisting of the elements $(x,y)$ for which, at
 a given value of $\Ph$, the QS-regime is possible. {\bf  Similar diagrams in terms of other variables
 can be found in \citep[][Fig. 2]{Christou2000} and \citep[][Fig. 6]{Mikkolaetal2006}.}

 Below we suppose to classify the qualitative properties of the asteroid's motion depending on the value $\xi$
 of the Hamiltonian $\Xi$. We will use the designation $D_{QS}(\xi,\Ph)$ to identify the subset of
 ${\cal B}_{QS}(\Ph)$ with elements for which there is a solution (11), corresponding to the motion
 in QS-orbit. It is easy to show that
$$
{\cal B}_{QS}(\Ph)=\cup_{\xi\ge\xi_{min}(\Ph)}D_{QS}(\xi,\Ph),
$$
 where $\xi_{min}(\Ph)$ denotes the minimal value of $\xi$, for which the motion in QS-orbit is
 possible for a given $\Ph$. We found numerically that $D_{QS}(\xi,\Ph)\rightarrow\partial
 {\cal D}(\Ph)$ as $\xi \searrow \xi_{min}(\Ph)$.

 Fig.~\ref{diagram1} demonstrates typical structural changes of $D_{QS}(\xi,\Ph)$, which take
 place as $\xi$ varies. If $\xi$ is slightly larger than $\xi_{min}(\Ph)$, the set
 $D_{QS}(\xi,\Ph)$ has a ring-like shape (Fig.~\ref{diagram1},b). At $\xi=\xi_h(\Ph)$,
a bifurcation takes place: holes emerge in the ring. As $\xi$ increases, the number of holes
 decreases from four (Fig.~\ref{diagram1},c) to two (Fig.~\ref{diagram1},d). Simultaneously,
 the size of the holes increases, and at $\xi=\xi_b(\Ph)$ their borders (the ``internal" borders
 of $D_{QS}(\xi,\Ph)$) approach the outer border of $D_{QS}(\xi,\Ph)$. The shapes of
 $D_{QS}(\xi,\Ph)$ for $\xi > \xi_b$ are shown in Fig.~\ref{diagram1},e and
 Fig.~\ref{diagram1},f. The difference of the last Figures stems from the existence of the
  bifurcation at $\xi=\xi_s(\Ph)$, when the triangular regions in the central part of
  ${\cal D}(\Ph)$ separate from the peripheral part of $D_{QS}(\xi,\Ph)$.

\begin{figure}[htb]
\vglue4.cm
\includegraphics[width=0.8\textwidth,keepaspectratio]{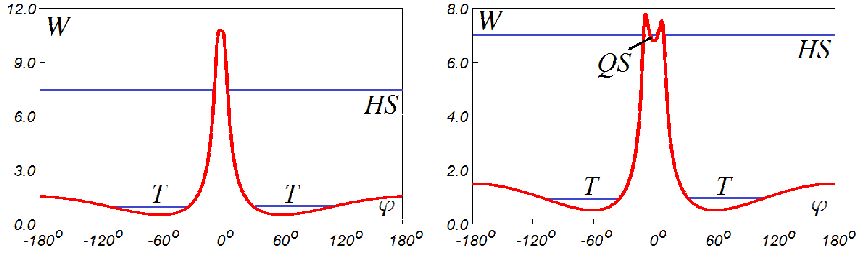}
\vglue-1.5cm

\caption{Behaviour of the averaged disturbing function $W$ at fixed $x,y,\Ph$.
Here $\Ph=0.99499$($\sigma=0.1$) in both cases,
$x=0.03898,y=-0.02251$ ($e=0.045,\omega=30^\circ,i=5.12871^\circ$) in the left panel,
$x=0.07368,y=-0.04254$ ($e=0.085,\omega=30^\circ,i=3.03062^\circ$) in right panel}
\label{Pic3}
\end{figure}

\begin{figure}[htb]
\vglue4.cm
\includegraphics[width=0.8\textwidth,keepaspectratio]{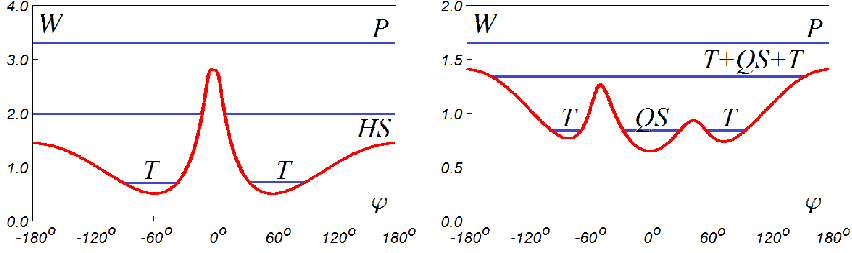}
\vglue-1.5cm

\caption{Behaviour of the averaged disturbing function $W$ at fixed $x,y,\Ph$.
Here $\Ph=0.86603$($\sigma=0.5$) in both cases,
$x=0.06500,y=-0.03753$ ($e=0.075,\omega=30^\circ,i=29.71851^\circ$) in the left panel,
$x=0.37711,y=-0.21772$ ($e=0.425,\omega=30^\circ,i=16.91652^\circ$) in right panel.}
\label{Pic4}
\end{figure}

Fig.~\ref{Pic3} and Fig.~\ref{Pic4} provide more examples of the behaviour of the function
$W$. Fig.~\ref{Pic3} illustrates the formation of a singularity of the
type $W\sim 1/\sigma$ at $\vphi=0$, as $\sigma \rightarrow 0$ (or $\Ph \rightarrow 1$).
In Section 4, this property will be used to analyze the asteroid motion in a QS-orbit with
 a small eccentricity and inclination.

The right panel in Figure~\ref{Pic4} demonstrates that at some $x,y,\Ph$  we can encounter
three different periodic solutions on the same level of the Hamiltonian of the ``fast"
subsystem.

In this paper, the analysis is limited to the case where
the behaviour of the function $W$ is described by graphs similar to
those presented in Fig.~\ref{Pic1} and Fig.~\ref{Pic3} (i.e. where a QS-orbit
shares the same energy level with a HS-orbit). The examination of the function
$W$ led us to conclude that this kind of behaviour is taking place for all
$(x,y)\in {\cal D}(\Ph)$ at $\Ph > P_h^*\approx 0.95924$ (or $\sigma < \sigma^*
\approx 0.28258$). As it follows from relations~(\ref{emax}) and~(\ref{imax}),
this corresponds to the motion of the asteroid in orbits
satisfying, for all moments of time, the inequalities
$$
e(t)<e_{max}^*\approx 0.28258,\quad
i(t)<i_{max}^*\approx 16.414^\circ.
$$
So the rest of the paper is devoted to an exploration of the properties of QS-orbits with small
and medium values of inclinations and eccentricities.

The general situation can be treated in the same way, but it will
be a rather cumbersome investigation.

\subsection{Averaging over the resonant phase variations. Evolutionary equations}

To study the long-term behaviour of slow variables, we use
evolutionary equations obtained by averaging of the right hand sides
of equations (\ref{SF}) over the solutions of the ``fast"
subsystem:

\beq{eveq} \frac{dx}{d\tau}=\eps \left\langle \frac{\partial
W}{\partial y} \right\rangle, \qquad \frac{dy}{d\tau}=-\eps
\left\langle \frac{\partial W}{\partial x}\right\rangle, \eeq
where
\beq{aver}
\left\langle \frac{\partial W}{\partial \zeta} \right\rangle =
\frac{1}{T(x,y,P_h,\xi)} \int_0^{T(x,y,P_h,\xi)}\frac{\partial
W}{\partial \zeta}(\vphi(\tau,x,y,P_h,\xi),x,y,P_h)d\tau,
\eeq
$$
\zeta=x,y.
$$

In the region $D_{QS}(\xi,\Ph)$, the averaging procedure
provides us with two vector fields, depending on what periodic solution
of the ``fast" subsystem was chosen in Eq. (\ref{aver}). One of them
describes the evolution of slow variables in the case of asteroid motion in a
QS-orbit, the other characterizes the evolution of an HS-orbit.

Applying the averaging procedure (\ref{aver}) we should take care of the situation
when the solution~(\ref{sol}) is non-periodic and corresponds
to the separatrix on the ``fast" subsystem's phase portrait. Taking this in mind
we define in the disk ${\cal D}(\Ph)$ the so-called uncertainty curve $\Gamma(\xi,\Ph)$
\citep{Wisdom1985,Neishtadt1987,NeishtadtandSidorenko2004,Sidorenko2006}:

\begin{center}
$\Gamma(\xi,\Ph)=\{(x,y)\in {\cal D}(\Ph), \exists
\vphi_*(x,y,\Ph)$ which is a point of a local or global maximum of the function $W(\vphi,x,y,\Ph)$
($x,y,\Ph$ being treated as parameters) satisfying the condition $W(\vphi_*,x,y,\Ph)=\xi \}$
\end{center}

\noindent As it is evident, for a given value of $\Ph$ the ``separatrix" solution exists on the level $\Xi=\xi$
only when $(x,y) \in \Gamma(\xi,\Ph)$.

\begin{figure}[htb]
\vglue4.cm
\includegraphics[width=0.8\textwidth,keepaspectratio]{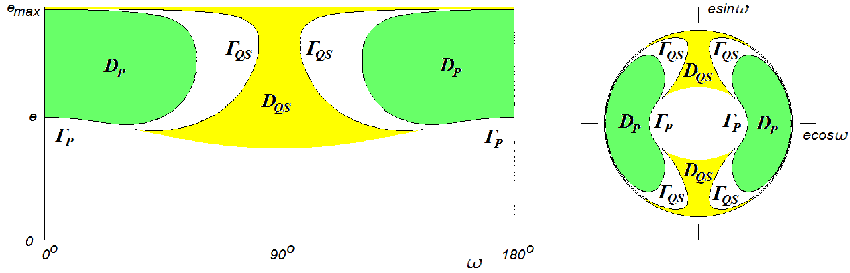}
\vglue-1.5cm

\caption{The location of the uncertainty curve components $\Gamma_{QS}$ and $\Gamma_P$
for $P_h=0.96825, \xi=3.5$. Green represents the areas of the asteroid motion in the
passing orbits at these values of $P_h$ and $\xi$.}
\label{curves}
\end{figure}

For $\xi \ge \xi_h(\Ph)$ the uncertainty curve consists typically of one or two
components: $\Gamma_P(\xi,\Ph)=\partial D_P(\xi,\Ph)$ and $\Gamma_{QS}(\xi,\Ph)\subset\partial D_{QS}(\xi,\Ph)$
(Fig.~\ref{curves}). For $\xi \in [\xi_{min}(\Ph),\xi_h(\Ph))$ it does not exist.

\begin{figure}[htb]
\vglue4.cm \hglue0.5cm
\includegraphics[width=0.75\textwidth,keepaspectratio]{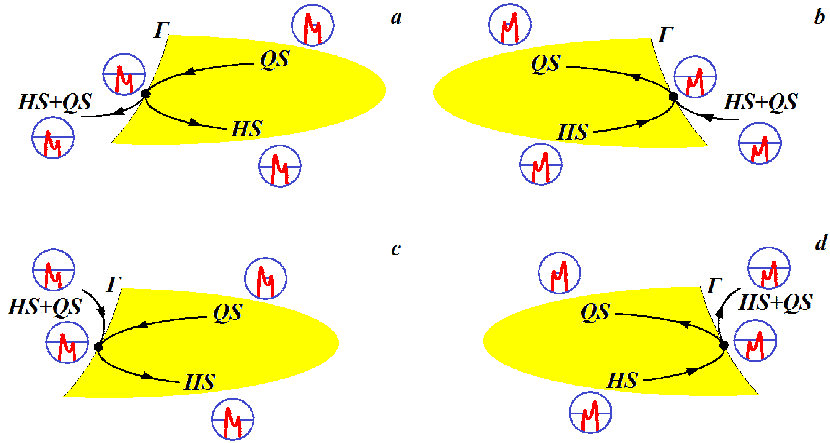}
\vglue-1.5cm

\caption{Matching of solutions to the evolutionary equations (\ref{eveq})
at the uncertainty curve $\Gamma(\xi,\Ph)$} \label{gluing}
\end{figure}

When the projection of the system (\ref{SF}) phase point on the plane $(x,y)$
approaches the uncertainty curve $\Gamma(\xi,\Ph)$ the qualitative changes in the behavior of the resonant phase
$\vphi$ take place (corresponding, for example, to a transition from a QS-orbit to an HS-orbit or back).
To prolongate the solutions of the averaged equations (\ref{eveq})
across the uncertainty curve, one can follow a rather straightforward strategy based
on the matching of solution with the same limit values at $\Gamma(\xi,\Ph)$.
In some cases, this matching is not unique. For example, in Fig.~\ref{gluing},a
we present a situation at the border of $D_{QS}(\xi,\Ph)$ when one
option for continuation results in exiting this region, with a further
motion in HS+QS-orbit, while the second option can be described as
reflection from the border, with a transition from QS-motion to HS-motion.
Since both options are possible in the system~(\ref{SF}), the evolution of
the motion near the uncertainty curve in this example has a probabilistic
nature\footnote{\bf If we consider the solution of the non-averaged system
with the initial conditions $x(0),y(0)$ exactly on $\Gamma$ the dynamics is definitely deterministic and depends
mainly on to what part of fast subsystem separatrix variables
$\varphi(0), \Phi(0)$ belong. But far from the uncertainty curve
(i.e., deep inside in the yellow zone in our
phase portraits) the initial conditions corresponding to
the crossing the above mention parts of the separatrix in the plane $(\varphi, \Phi)$ are mixed strongly.
And a small uncertainty in the initial conditions does not allow us to predict uniquely the
qualitative behavior of the system when the phase point leave the vicinity of $\Gamma$.
Nevertheless, for the set of the possible initial conditions
we can compare the measures of subsets resulting in the different qualitative behavior later on and then characterize the
dynamics in the probabilistic way.} \citep{Arnoldetal2006}.
Similar example is provided by Fig.~\ref{gluing},d. Fig.~\ref{gluing},b
and Fig.~\ref{gluing},c present situations where the solution leaves
the neighbourhood of $\Gamma_{QS}(\xi,\Ph)$ in a unique way.

\subsection{Adiabatic approximation}

{\bf The evolutionary equations (\ref{eveq}) provide so called ``adiabatic approximation"
of the system's long-term dynamics \citep{Wisdom1985, Yokoyama1996}. Indeed the first
two equations in (\ref{SF}) correspond to 1DOF Hamiltonian system depending on slowly
varying parameters $x,y$. Therefore this system approximately preserves the value of the adiabatic
invariant \citep{Henrard1982,Arnoldetal2006}
\beq{ai}
I(x,y,P_h,\xi)=\oint \Phi\,d\vphi.
\eeq

For averaged equations (\ref{eveq}) the adiabatic invariant (\ref{ai}) is an exact integral.
So the phase trajectories $x(\tau),y(\tau)$ are laying on its level curves. The matching of the
phase trajectories at $\Gamma$ satisfies the conditions
$$
I_{HS+QS}=I_{QS}+I_{HS}
$$
and
$$
2I_{P}=I_{HS+QS}\quad \mbox{or}\quad 2I_{P}=I_{HS}
$$
for the components $\Gamma_{QS}$ and $\Gamma_P$ respectively (indices denote the orbital regimes used
to compute the adiabatic invariant (\ref{ai})).

The validity of the adiabatic approximation is limited due the violation of the adiabatic
invariance in the vicinity of the uncertainty curve $\Gamma$. The related
phenomena are discussed in Sec. 3.2.}

\section{Investigation of secular effects on the basis of evolutionary equations }

\subsection{Comparative analysis for different level sets of the
 Hamiltonian $\Xi$}

 A good way to understand the evolution of slow variables is to consider the phase portraits
 of system (\ref{eveq}). {\bf As it follows from Sec. 2.5 the drawing of the phase portraits
 reduces to the drawing of the level curves of the adiabatic invariants $I_{P,QS,\ldots}(x,y,P_h,\xi)$
 for given values $P_h,\xi$.}

\begin{figure}[htb]
\vglue3.7cm \hglue-.5cm
\includegraphics[width=0.85\textwidth,keepaspectratio]{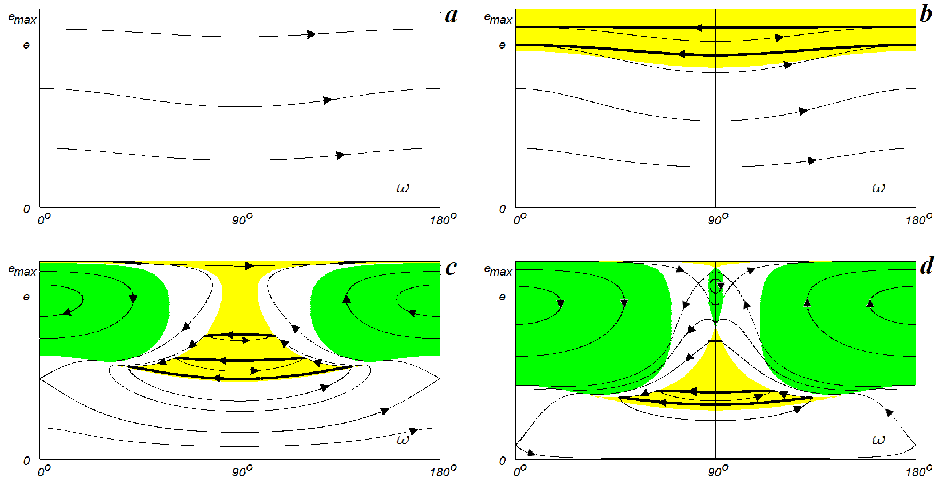}
\vglue-1.5cm 

\caption{Examples of the secular evolution at 1:1 mean motion resonance.
Thick lines correspond to motions in QS-orbits. Two sets of phase trajectories in the domain $D_{QS}(\xi,\Ph)$
(painted yellow) demonstrate the
dependence of secular effects on the character of motion (HS-orbit or QS-orbit) on
intermediate time scale. In all cases $\Ph=0.96825$ ($e_{max}=0.25$). The values of
Hamiltonian of SF-system (\ref{SF}): a - $\xi=1.75$, b - $\xi=2.35$, c - $\xi=3.5$,
d - $\xi=4.5$} \label{pp}
\end{figure}

 Several examples of phase portraits are given in Fig.~\ref{pp}. For better visualization,
 these portraits present the behaviour of the averaged osculating elements $\omega$ and $e$
 given by formulae~(\ref{oe}) and~(\ref{om}). Dependent on the values of $\xi,P_h$, these phase portraits differ
 in the number of equilibrium points and/or the separatrices placement. There exist some
 other important features which depend on $\xi,P_h$ also.

 {\bf It is interesting to compare these phase portraits with the evolutionary diagrams obtained in~\citep{Namouni1999} by
 means of direct integration of motion equations.} The phase portrait in Fig.~\ref{pp} resembles
 Fig. 12 from \citep{Namouni1999} and represents the situation when the motion in a QS-orbit
 is impossible ($\xi < \xi_{min}(\Ph)$). All trajectories are related to HS-orbits with a
 circulating argument of the pericenter, $\omega$.

 The relatively simple portrait in Fig.~\ref{pp},b is typical for $\xi \in (\xi_{min}, \xi_h)$.
 It is the case when transitions between the motion in a QS-orbit and that in an HS-orbit are
 impossible. As a remarkable property of this case, we can mention the opposite directions of the pericenter
 circulation for these orbits. {\bf There is no similar diagram in~\citep{Namouni1999}, because the motion
 equations were integrated only with the initial conditions corresponding to P-orbits and HS-orbits.}

 More complex behaviour of phase trajectories is presented on the phase portrait in
 Fig.~\ref{pp},c (compare with Fig. 13 in~\citep{Namouni1999}). One of the most interesting properties
 of asteroid motion is associated with the closed contours composed by the phase trajectories
 related to a QS-orbit and an
 HS-orbit. Such a contour corresponds to the alternation of the QS- and HS-regimes in orbital
 motion over a large enough time span. Nevertheless, at a certain moment, this alternation
 can be violated due to the escape from $D_{QS}(\xi,\Ph)$. The opportunity of escape is
 provided by outgoing trajectory matched to the described contour
(Fig.~\ref{gluing},a or d). Numerically, this phenomenon was established by~\citet{Namouni1999}.

The phase portrait in Fig.~\ref{pp},d demonstrates the further complication of
asteroid dynamics as $\xi$ increases. The fission of $D_{QS}(\xi,\Ph)$ (the
separation of triangular areas from peripheral part) at $\xi=\xi_s$ is accompanied
by the appearance of a stable equilibrium corresponding to motion in passing orbit
with "frozen" pericenter: $\omega=90^\circ$ or $\omega=270^\circ$ (Lidov-Kozai resonance).
{\bf In \citep{Namouni1999} the passing orbits with $\omega$ librating around Lidov-Kozai resonance
at $\omega=90^\circ$ or $\omega=270^\circ$ can be seen in Fig. 18.}

\begin{figure}[htb]
\vglue3.5cm \hglue1.cm
\includegraphics[width=0.70\textwidth,keepaspectratio]{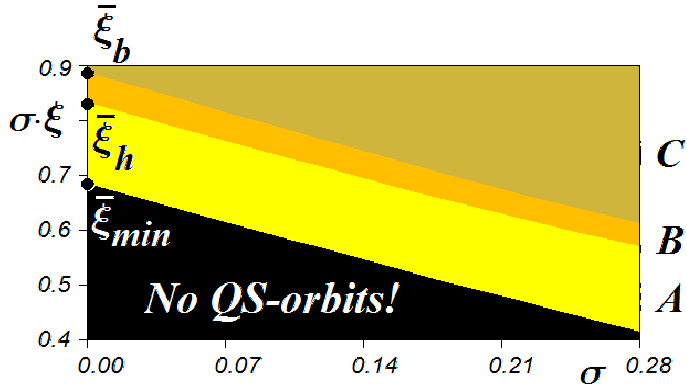}
\vglue-1.5cm

\caption{Dependence of QS-orbit properties on parameters $\sigma,\xi$.
Area A: any motion in QS-orbit preserves
its character forever. Area B: due to the holes in $D_{QS}(\xi,\Ph)$
(e.g., Fig.3,c and Fig.3,d) for some initial values the alternating escapes from
QS-orbits and returns back become possible. Area C: the configuration
of $D_{QS}(\xi,\Ph)$ (e.g., Fig.3,e and Fig.3,f) allows permanent motion in QS-orbit
only at small inclination} \label{cond}
\end{figure}

In Fig.~\ref{cond} we tried to summarize some information about the motion
in QS-orbits. The choice of parameters ($\sigma$ and $\xid=\xi\cdot\sigma$) is justified
by our intention to simplify the structure of the presented diagram. As one can see,
there are practically straight-line borders between regions with different properties
of QS-orbits, and the following limits exist:
$$
\xid_{min}=\lim_{\sigma \rightarrow 0}\sigma\cdot\xi_{min}(\Ph)=0.68644...,\quad
\xid_h=\lim_{\sigma \rightarrow 0}\sigma\cdot\xi_h(\Ph)=0.8346...,
$$
$$
\xid_b=\lim_{\sigma \rightarrow 0}\sigma\cdot\xi_b(\Ph)=0.886...
$$
The regularity of the system properties with respect to proper scaling of parameters
stimulates an application of the appropriate perturbation technique for analysis
of asteroid dynamics in case $\sigma \ll 1$ (Section 4).

\subsection{Chaotization}

Matching of phase trajectories on the uncertainty curve $\Gamma(\xi,\Ph)$ provides ``zero-order" theory
of slow variables evolution in the case of qualitative transformation of fast
variables behaviour. Actually in the neighbourhood of $\Gamma(\xi,\Ph)$ the adiabatic invariance
is violated. As a consequence the projection
of a phase point of the system~(\ref{SF}) onto the plane $x,y$ jumps in quasi-random
way from the incoming trajectory of the averaged system~(\ref{eveq}) to some outgoing trajectory
 located at the distance of the order $\eps|\ln \eps|$ from the outgoing trajectory obtained by a
formal matching. The accurate estimations of the quasi-random jumps were given
in~\citep{Timofeev1978, Caryetal1986, Neishtadt1986, Neishtadt1987b}.

This mechanism of chaotization (giving rise to the so-called adiabatic chaos
in the asteroid's dynamics) is typical for mean-motion
resonances~\citep{Wisdom1985,Neishtadt1987,NeishtadtandSidorenko2004,
Sidlichovsky2005,Sidorenko2006,BatyginandMorbidelli2013}. {\bf It is likely the responsible
for the increment of the eccentricity observed by~\citet{Namouni1999} in the
series of the transitions $QS\rightarrow HS \rightarrow \ldots$.}

 \section{Dynamics of the asteroid moving in an orbit of a small eccentricity and inclination}

 In the case of the asteroid moving in an orbit with a small eccentricity and inclination,
 the resonance condition
\begin{center}
$|n - n'|$\raisebox{2pt}{$\mathop{<}\limits_{\displaystyle \sim}$}$\mu^{1/2}$
\end{center}
implies $\Ph \approx 1$ or $\sigma=\sqrt{1-\Ph^2}\ll 1$ (Section 2.1). {\bf Using different
simplifying assumptions \citet{Namouni1999} derived analytical expressions characterizing
the secular evolution of QS-orbits, HS-orbits and P-orbits.

Our goal is to provide a qualitative description of long-term dynamics for any resonant orbit
with a small eccentricity and inclination satisfying the only restriction imposed at the beginning
of the paper: do not approach the Hill sphere of the planet. In particular it implies
$\mu^{1/3}\ll \sigma \ll 1$. To achieve our goal we must consider the motions which do not
satisfy the simplifying assumptions applied in  \citep{Namouni1999}.}

 \subsection{The leading term in the expression for the averaged disturbing function $W$}

 Graphs presented in Fig.~\ref{Pic3} demonstrate two important properties of the averaged
 disturbing function at $\sigma \to 0$ ($\Ph \to 1$): ~(i) the increase of its values for
 a resonant phase $\vphi$ close to zero, and (ii) the decrease of the profile ``thickness"
 (i.e., the decrease of the interval of resonant phase values at which $W\sim 1/\sigma$).
 Analytically, it implies the following structure for the function $W$:

\beq{WS}
W(\vphi,x,y,\sigma)=\frac{1}{\sigma}\Wd\left(\frac{\vphi}{\sigma},
\frac{x}{\sigma},\frac{y}{\sigma}\right)+\{\mbox{Residual term which is in general of the order of 1}\},
\eeq
where $\vphi\in[-\pi,\pi]$ and
$$
\Wd(\vphid,\xd,\yd)=\frac{1}{2\pi}\int^{\pi}_{-\pi}\frac{d\hd}{\Delta},
$$
$$
\Delta(\vphid,\xd,\yd,\hd)=\lim_{\sigma\to 0}
\left.\frac{1}{\sigma}|\rb(\vphi,x,y,\hd)-\rb(\lambda')|
\right|_{\vphi=\sigma \vphid,\,x=\sigma \xd,\,y=\sigma \yd}=
$$
$$
\vphid^2 - 4\vphid(\xd \sin \hd - \yd \cos \hd)+ 3(\xd \sin \hd - \yd \cos \hd)^2+
$$
$$
(\xd^2 + \yd^2) + [1 - (\xd^2 + \yd^2)]\sin^2 \hd.
$$

 Now we shall address the approximate evolutionary equations obtained by approximating $W$ with
 its leading term only. To that end, it will be useful to discuss the main properties of the
 function $\Wd(\vphid,\xd,\yd)$.

 The function $\Wd$ is defined on the set $R^1\times {\cal D}^2\setminus{\cal D}_0$, where
$$
{\cal D}^2 = \{(x,y)\in R^2|\,x^2+y^2\le1\},\quad
{\cal D}_0=\{\vphid,\xd,\yd|\,\exists \hd:\Delta(\vphid,\xd,\yd,\hd)=0\}.
$$
 The following symmetries are easy to verify:
$$
\Wd(\vphid,\xd,\yd)=\Wd(\vphid,-\xd,\yd)=\Wd(-\vphid,\xd,-\yd)=\Wd(-\vphid,\xd,\yd).
$$

 It is noteworthy that, for a given values $\xd,\yd$, the function $\Wd$ is even with respect
 to $\vphid$, while the function $W$ does not, in general, possesses this property. In the
 studies of QS-orbits, the excessive symmetry of the dynamical model based on truncated
 expression for the disturbing function were mentioned previously by~\citet{Mikkolaetal2006}. This symmetry should not be surprising: in a non-explicit way, here
 we apply the Hill's approximation of the three body problem; the excessive symmetry of
 Hill's approximation (in comparison with the original problem) is well known~\citep{Henon1997}.

\begin{figure}[htb]
\vglue-6cm \hglue3.cm
\includegraphics[width=\textwidth,keepaspectratio]{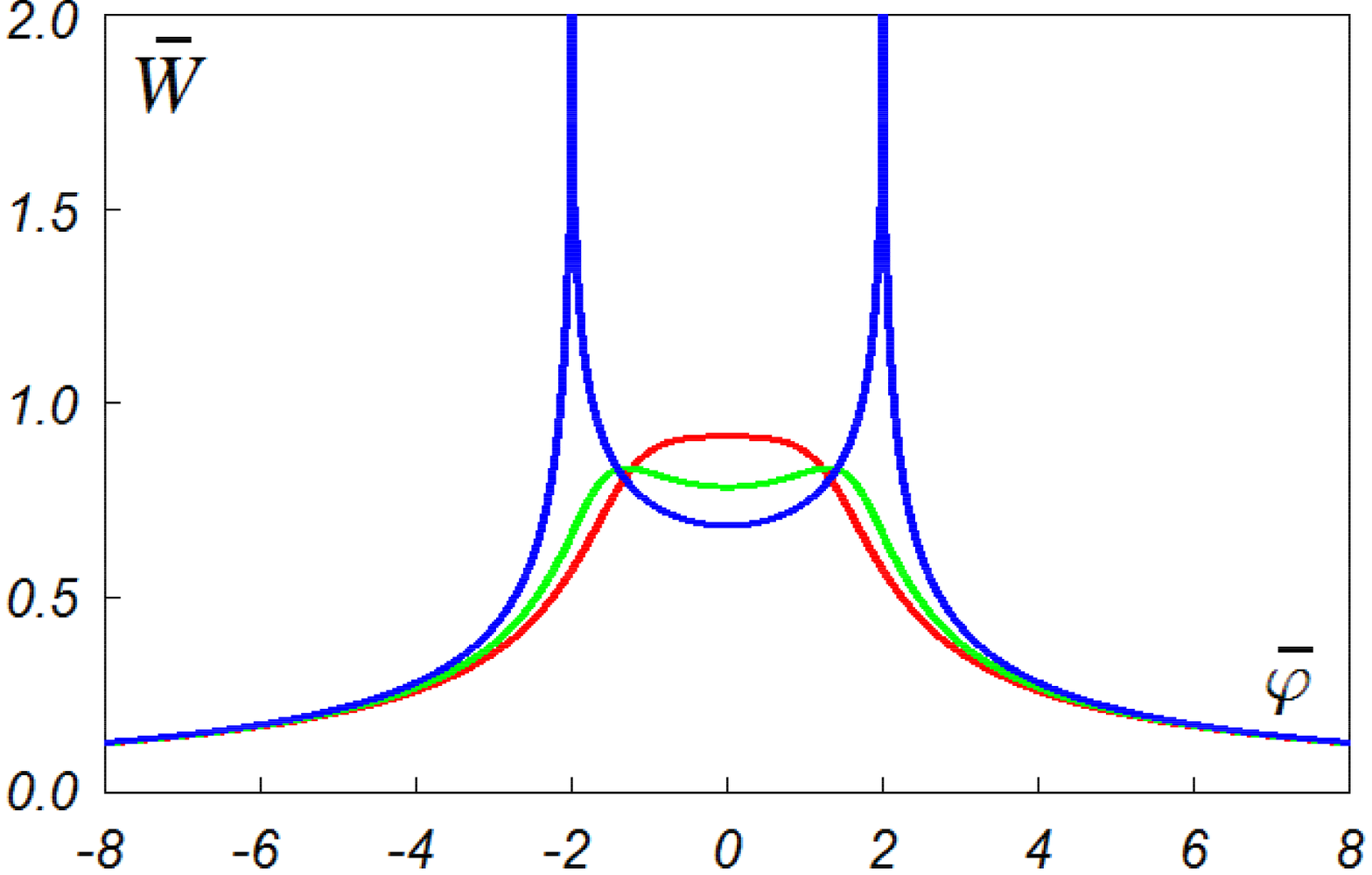}
\vglue5.5cm

\caption{The behaviour of the function $\Wd$ at some
fixed values of $\overline{x},\overline{y}$:
$\overline{x}=0.7,\,\overline{y}=0.0$(red line);
$\overline{x}=0.85,\,\overline{y}=0.0$(green line);
$\overline{x}=1.0,\,\overline{y}=0.0$(blue line)} \label{W3}
\end{figure}

 The function $\Wd(\vphid,\xd,\yd)$ and its derivatives
$$
\frac{\partial \Wd}{\partial \xd},\quad
\frac{\partial \Wd}{\partial \yd},\quad
\frac{\partial \Wd}{\partial \vphid}
$$
 can be expressed in terms of the elliptic integrals of the first and second kind (Appendix B).
 One can use these expressions to accelerate numerical computations
 at the stage of averaging over the resonant phase oscillations/rotations.

 In the case of
$$
\vphid \sim (\xd^2+\yd^2)^{1/2} \ll 1~,
$$
 the following approximate formula can be derived:
\beq{WDA}
\Wd(\vphid,\xd,\yd)\approx \frac{1}{2\pi}
\ln \frac{256}{\left[\vphid^2 + (\xd^2+4\yd^2)\right]^2-16\vphid^2\yd^2}~.
\eeq
 Expression (\ref{WDA}) allows one to understand better the behaviour of the function
 $\Wd(\vphid,\xd,\yd)$ for small enough $\vphid,\xd,\yd$, i.e., when this function becomes
 singular.

\begin{figure}[htb]
\vglue-6cm \hglue3.0cm
\includegraphics[width=\textwidth,keepaspectratio]{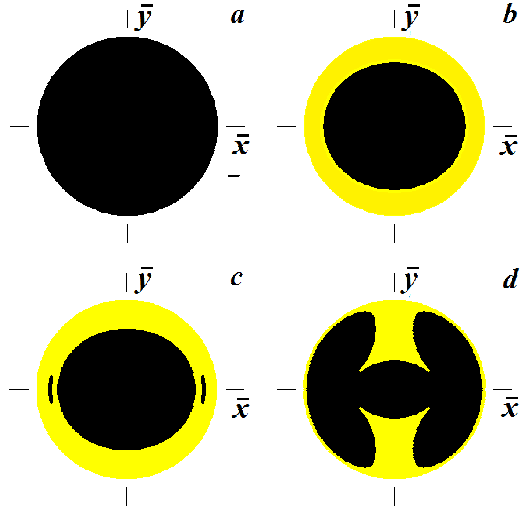}
\vglue3.5cm

\caption{Examples of the set $\overline{D}_{QS}(\overline{\xi})$ structure:
a - $\overline{\xi} < \overline{\xi}_{min}$
($\overline{D}_{QS}(\overline{\xi})=\oslash$), b - $\overline{\xi}=0.82 \in
(\overline{\xi}_{min},\overline{\xi}_h)$,
c - $\overline{\xi}=0.84 \in (\overline{\xi}_h,\overline{\xi}_b^*)$,
d - $\overline{\xi}=1.2 > \overline{\xi}_b^*$ }\label{Diagram2}
\end{figure}

 \subsection{Derivation of the evolutionary equations
 in the case of the leading term approximation for $W$}

If the leading term approximation is used for $W$, the equations
for fast subsystem (\ref{FS}) can be rewritten as
\beq{FSD}
\frac{d\vphid}{d\taud}= 3 \Phid, \quad
\frac{d\Phid}{d\taud} = - \frac{\partial \Wd}{\partial \vphid}~, \eeq
where
$$
\taud = \sigma^{-3/2}\tau, \; \Phid=\sigma^{1/2}\Phi, \; \vphid=\sigma^{-1}\vphi.
$$

 We are interested in oscillatory solutions to Eqs.~(\ref{FSD}), as this naturally provides
 the approximation for solutions to Eqs.~(\ref{FS}), corresponding to QS-regimes of the
 orbital motion. Let the pair
\beq{osd}
\vphid(\taud,\xd,\yd,\xid)\,,\;\Phid(\taud,\xd,\yd,\xid)\,,
\eeq
denote a solution to Eqs.~(\ref{FSD}), satisfying the equality
$$
\frac{3}{2}\Phid^2(\taud,\xd,\yd,\xid)+\Wd(\vphid(\taud,\xd,\yd,\xid),\xd,\yd)=\xid
$$
at any moment $\taud \in R^1$ ($x,y$ being considered as fixed parameters). This solution
exists only in the case of $(x,y)\in \Dd_{QS}(\xid)$, where
$$
\Dd_{QS}(\xid)=\left\{(x,y)\in D_2: \Wd(0,\xd,\yd)<\xid,\;
\frac{\partial^2 \Wd}{\partial \vphid^2}(0,\xd,\yd)>0\right\}
$$
 Fig.~\ref{Diagram2} illustrates the dependence of $\Dd_{QS}(\xid)$ on $\xid$. The changes
 in the structure of this set resemble the changes of the set $D_{QS}(\xi,\Ph)$ when $\xi$
 varies (Fig.~\ref{diagram1}), although in general the leading term approximation of the disturbing
 function results in the loss of certain fine details (some of them we discuss below). For $\xid \in (\xid_{min},\xid_h)$, the set $\Dd_{QS}(\xid)$
 has a ring-like structure ($\xid_{min},\xid_h$ have been defined at the
 end of Section 3.1). Using the expression for $\Wd$ in terms of elliptic integrals,
 one obtains:
$$
\xi_{min}=\frac{1}{\pi}K\left(\sqrt{\frac{3}{4}}\right)
$$
 At $\xid=\xid_h$, holes emerge in $\Dd_{QS}(\xid)$; at $\xid=\xid_b^*$
($\xid_b^*=0.89542...$)
the borders of these holes approach the ``external" border of
$\Dd_{QS}(\xid)$. The difference between the values of $\xid_b$ and $\xid_b^*$
is probably the consequence of the  excessive symmetry of Hill's approximation mentioned above
(one more consequence being the existence of only two holes
in $\Dd_{QS}(\xid)$ for all $\xid \in (\xid_h,\xid_b^*)$).

 The solution~(\ref{osd}) is used to obtain an approximate expression for the right-hand
 sides of equations (\ref{eveq}) describing the evolution of the slow variables on the
 level set $\Xi=\xid/\sigma$ for the case of the asteroid moving in a QS-orbit:
\beq{ARP}
\left\langle \frac{\partial W}{\partial \zeta}\right\rangle \approx \frac{1}{T(\xid/\sigma,\Ph(\sigma))}\int_0^{T(\xid/\sigma,\Ph(\sigma))}
\frac{1}{\sigma^2}\frac{\partial \Wd}{\partial \zetad}
\left(\frac{\vphi}{\sigma}\left(\tau,\Ph(\sigma),\frac{\xid}{\sigma}\right),
\frac{x}{\sigma},\frac{y}{\sigma}\right)d\tau \approx
\eeq
$$
\approx
\frac{1}{\sigma^2 \Td(\xid)} \int_0^{\Td(\xid)}\frac{\partial \Wd}{\partial \zetad}
\left(\vphid(\taud,\xid),\frac{x}{\sigma},\frac{y}{\sigma}\right)d\taud.
$$
Here $\Td(\xid)$ denotes the period of oscillatory solutions~(\ref{osd}), while
$\zeta=x,y$ and $\zetad=\xd,\yd$.

 In the case of the asteroid moving in  HS- or P-orbits, construction of similar
 approximate expression is a more delicate procedure since there is no suitable periodic
 solution to the auxiliary system (\ref{FSD}). However one can use non-periodic
 solutions to this system. Specifically, let us consider the solution to Eqs~(\ref{FS}), corresponding to motion
 in an HS-orbit with $\Xi=\xid/\sigma \gg 1$. A crude (but sufficient for our purposes)
 estimate of its period is
\beq{TA}
T\approx \pi \sqrt{\frac{8\sigma}{3\xid}}~.
\eeq
 Due to the afore-mentioned symmetry of the leading term in the expression for the disturbing
 function, averaging can be performed over the reduced time interval:
\beq{ARP0}
\left\langle \frac{\partial W}{\partial \zeta}\right\rangle \approx
\frac{4}{T}\int_0^{T/4}\frac{1}{\sigma^2}\frac{\partial \Wd}{\partial \zetad}
\left(\frac{\vphi}{\sigma}\left(\tau,\Ph(\sigma),\frac{\xid}{\sigma}\right),
\frac{x}{\sigma},\frac{y}{\sigma}\right)d\tau
\eeq
 Then we can replace $\vphi(\tau,\xid/\sigma,\Ph(\sigma)/\sigma)$ in the expression (\ref{ARP0}) with the
 function $\vphid(\taud,\xid)$, which provides (in combination with $\Phid(\taud,\xid)=
 d\vphid/d\taud$) a non-periodic solution to (\ref{FSD}), satisfying the conditions:
$$
\Wd\left(\vphid(0,\xid),\frac{x}{\sigma},\frac{y}{\sigma}\right)=\xid,\;
\vphid(\taud,\xid) \rightarrow +\infty\;\mbox{at}\; \taud\rightarrow +\infty~.
$$
 This results in
\beq{ARP1}
\left\langle \frac{\partial W}{\partial \zeta}\right\rangle \approx
\frac{4}{T}\int_0^{\sigma^{-3/2}T/4}\frac{1}{\sigma^{1/2}}\frac{\partial \Wd}{\partial \zetad}
\left(\vphid(\tau,\xid),
\frac{x}{\sigma},\frac{y}{\sigma}\right)d\taud~.
\eeq
 To finish up, we replace the period $T$ in Eq. (\ref{ARP1}) with
 its approximate expression (\ref{TA}) and expand the integration over the positive semiaxis
 (since $\sigma^{-3/2}T(\xid/\sigma)$ approaches infinity as $\sigma\rightarrow 0$):
\beq{ARP2}
\left\langle \frac{\partial W}{\partial \zeta}\right\rangle \approx
\frac{1}{\pi\sigma}\sqrt{23\xid}
\int_0^{+\infty}\frac{\partial \Wd}{\partial \zetad}
\left(\vphid(\taud,\xid),\frac{x}{\sigma},\frac{y}{\sigma}\right)d\taud.
\eeq
 The convergence of the integral in Eq.~(\ref{ARP2}) can be proven rigorously: it follows from
 the rapid decrease of the function $\Wd$ at $\vphid \rightarrow \infty$. Physically,
 the approximation (\ref{ARP2}) means that in the case of motion in an HS-orbit
 the evolution of slow variables ($e,\omega,i,\Omega$) is substantial only
 during close approaches of the asteroid to the planet.

 In a similar way, the ``averaging" can be done for the motion in a P-orbit.
 Formally, the ensuing expression coincides with (\ref{ARP2}), but the function
 $\vphid(\taud,\xid)$ is rendered in this case by a solution to Eq.~(\ref{FSD}),
 satisfying the conditions:
 $$
 \vphid(0,\xid)=0,\;
 \vphid(\taud,\xid) \rightarrow +\infty\;\mbox{at}\; \taud\rightarrow +\infty~.
 $$

\subsection{Dynamics of the asteroid. Estimations based on the approximate
evolutionary equations}

 Insertion of expressions (\ref{ARP}) and (\ref{ARP2}) into Eqs.~(\ref{eveq}) furnishes approximate
 equations characterising the secular evolution of asteroid motion in a near-circular low-inclination
 orbit in the 1:1 resonance.

\begin{figure}[htb]
\vglue3.0cm \hglue2.0cm
\includegraphics[width=0.60\textwidth,keepaspectratio]{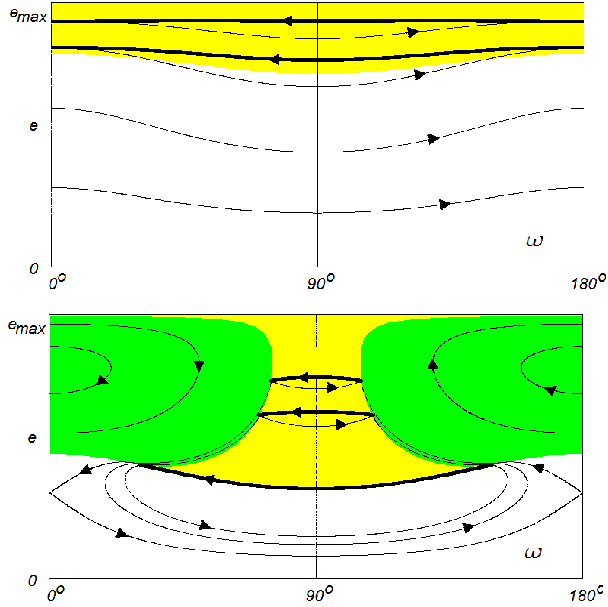}
\vglue-0.5cm
\caption{Examples of phase portraits obtained in the leading term
approximation for $W$: a - $\xi/\sigma=0.82<\overline{\xi}_h$,
b - $\xi/\sigma=1.2 > \overline{\xi}_b$} \label{hpp}
\end{figure}

 In Fig.~\ref{hpp}, we present examples of the phase portraits corresponding to the approximate
 evolutionary equations. The behaviour of the phase trajectories has a remarkable similarity:
 it does not depend on $\sigma$, if we use $e_{max}$ as a unit for length along the vertical
 axis. Fig.~\ref{hpp},a is practically identical to Fig.~\ref{pp},b -- it illustrates
 that the approximate equations are best of all suited for the description of the motion without qualitative changes
 in the behavior of the resonant phase $\vphi$ (i.e., asteroid moves permanently in a
 QS- or HS-orbit).

 Fig.~\ref{hpp},b resembles Fig.~\ref{pp},c except for the location of the regions
 corresponding to P-orbits (these regions are painted in green). The difference emerges due
 to the absence of compound QS+HS-regimes in the model based on the leading-term approximation
 of $W$ (this is another consequence of its excessive symmetry). So the model discussed above should be
 applied with caution for the analysis of motions with transitions between different types of
 orbits.

The approximate expressions (\ref{ARP}) and (\ref{ARP2}) for the averaged
derivatives of $W$ allow one to estimate the rates of the evolution of the orbital elements:
\beq{e1}
\frac{de}{dt},\;\frac{di}{dt}\sim\frac{\mu}{\sigma^2},\;
\frac{d\Omega}{dt},\;\frac{d\omega}{dt}\sim\frac{\mu}{\sigma^3}
\eeq
for the case of motion in a QS-orbit, and
\beq{e2}
\frac{de}{dt},\;\frac{di}{dt}\sim\frac{\mu}{\sigma},\;
\frac{d\Omega}{dt},\;\frac{d\omega}{dt}\sim\frac{\mu}{\sigma^2}~,
\eeq
for other types of orbital motion. As it follows from Eqs.~(\ref{e1}),(\ref{e2}),
evolution is faster for motion in a QS-orbit. This is not surprising since
for such an orbit the mean distance from the asteroid to the planet is smaller than
for other types of orbits and, consequently, the perturbations due to the attraction of the asteroid
by the planet are more substantial.

For motion with transitions between different types of orbital behaviour, formulae
(\ref{e1}) and (\ref{e2}) yield the following estimates
\beq{e3}
T_{QS}\sim\frac{\sigma^3}{\mu},\quad
T_{HS,P,\ldots}\sim\frac{\sigma^2}{\mu},
\eeq
where $T_{QS}$ and $T_{HS,P,\ldots}$ denote the duration of motion in a (temporary)
QS-orbit and in orbits of other types, respectively. An evident consequence of (\ref{e3})
is:
$$
T_{QS}\ll T_{HS,P,\ldots}~.
$$

{\bf As it was mention at the beginning of this Section under some additional assumptions
\citet{Namouni1999} derived analytical expressions
describing the secular evolution of the orbits under investigation. In particular, his formulas for
QS-orbits are limited to the case when the resonant phase $\varphi$ does not oscillate - i.e., the
``fast" subsystem is in ``quasi-steady" equilibrium close to local minimum of the averaged disturbing
function $W(\varphi,x,y,P_h)$ (in our phase portraits the motions in these orbits correspond to
trajectories at the lower boundary of the yellow areas). The HS-orbits in \citep[][Sec. 4.1]{Namouni1999}
approach the close vicinity of the planet's Hill sphere what is beyond the scope of our analysis.
The consideration of P-orbits in \citep[][Sec. 4.2]{Namouni1999} will be commented in the Section 4.4.}

\subsection{\bf Some remarks about the properties of the motion in the passing orbits
with small eccentricity and inclination}

{\bf The relation of the Fig.~\ref{Pic_cc} and Fig.~\ref{pp},d leads to a conclusion
that there are two topologically different types of the passing orbits. The passing orbits
with $\omega$ librating around $0^\circ$ or $180^\circ$ are linked with the orbit of the
planet. The libration of $\omega$ around $90^\circ$ or $270^\circ$ takes place in the case
of the motion in unlinked passing orbits.

To study the motion in linked passing orbits \citet{Namouni1999} applied the averaging
over the resonant phase $\varphi$ neglecting the variations in its rate. Formally it is
identical to the averaging over motion in the close non-resonant orbits undertaken
by \citet{LidovandZiglin1974}.
As a result, the expressions for double-averaged disturbing function in these papers differ
by designations only. But then \citet{LidovandZiglin1974} only mentioned the existence of
the ``centre" type equilibrium at $e=\sigma/\sqrt{2}$, $\omega=0^\circ$ or $\omega=180^\circ$,
while \citet{Namouni1999} derived the analytical solution to the system of the evolutionary
equations.

For small but non-zero values of $\sigma$ the regions of the unlinked orbits can be described
as the tiny triangles encompassing the lines $\omega=90^\circ$ and $\omega=270^\circ$.
As a consequence of the results presented in \citep{LidovandZiglin1974} one can obtain for
unlinked passing orbits
the existence of the ``centre" type equilibrium at $e\approx\sigma\sqrt{\frac{2}{3}}$, $\omega=90^\circ$
or $\omega=270^\circ$, what is in a good agreement with the numerical investigations.}

\section{Example: the future escape of asteroid 2004GU9 from the QS-orbit}

 Now we apply the evolutionary equations~(\ref{eveq}) for the analysis of secular effects in
 orbital motion of the near-Earth asteroid 2004GU9. The restricted circular
 three-body problem is insufficient for accurate investigation of the actual asteroid dynamics.
 So we are interested more in understanding of the time scales involved and of some other
 quantitative characteristics of the phenomena discussed in previous Sections.

 Currently, asteroid 2004GU9 is moving along a QS-orbit with osculating elements presented in
 Table 1~\citep{Mikkolaetal2006,Wajer2010}. We chose it among the other quasi-satellites of the Earth
 due to the absence of this object's close encounters with Venus and Mars -- this restriction
 justifies, to some extent, the investigation of the secular effects under the scope of RC3BP.

\begin{figure}[htb]
\vglue4.cm \hglue1.5cm
\includegraphics[width=0.6\textwidth,keepaspectratio]{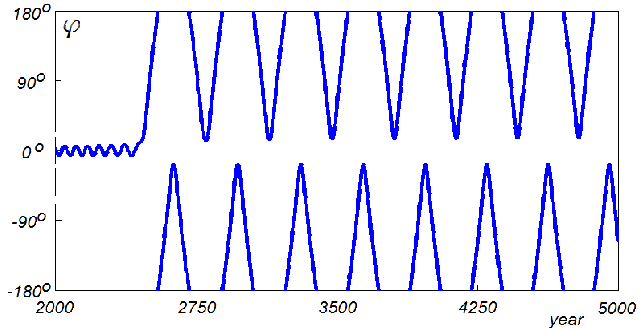}
\vglue-1.5cm
\caption{The behaviour of the resonant phase $\vphi$ of 2004GU9} \label{aph}
\end{figure}

 Fig.~\ref{aph} demonstrates the behaviour of the resonant phase $\vphi$ according to the
 results of direct numerical integration of the equations of motion corresponding to RC3BP
 with the initial values provided by the elements in Table~1 and the mass parameter $\mu=3.04
 \cdot10^{-6}$ (we added the mass of the Moon to the mass of the Earth). The
 motion in a QS-orbit will last for approximately 500 years, with a subsequent transition
 to an HS-orbit. In~\citep{Wajer2010}, similar conclusion has been achieved via taking into account
 the gravitational pull of all other Solar system planets, as well as the Moon and Pluto.

\noindent
\begin{center}
{\bf Table 1} Osculating orbital elements of asteroid 2004GU9.

Epoch: March 14, 2012 (JD2456000.5)

\begin{tabular}{cc}
\hline
\makebox[4cm][c]{Element} & Value\rule{0pt}{15pt}\\
\hline
$a(AU)$                     & $1.001056350821795$  \rule{0pt}{15pt}\\
$e$                         & $.1362904920360489$  \rule{0pt}{15pt}\\
$i(\vphantom{h}^\circ)$      & $13.64944749947083$  \rule{0pt}{15pt}\\
$\Omega(\vphantom{h}^\circ)$ & $38.74489028357296$  \rule{0pt}{15pt}\\
$\omega(\vphantom{h}^\circ)$ & $280.6255989836612$  \rule{0pt}{15pt}\\
$M(\vphantom{h}^\circ)$      & $217.2153150601352$  \rule[-7pt]{0pt}{22pt}\\
\hline
\end{tabular}
\vskip.1in

{\it Note.} The values of the orbital elements were borrowed from the JPL Small-Body Database
\end{center}
\vskip.1in

\begin{figure}[htb]
\vglue4.cm \hglue-.5cm
\includegraphics[width=0.85\textwidth,keepaspectratio]{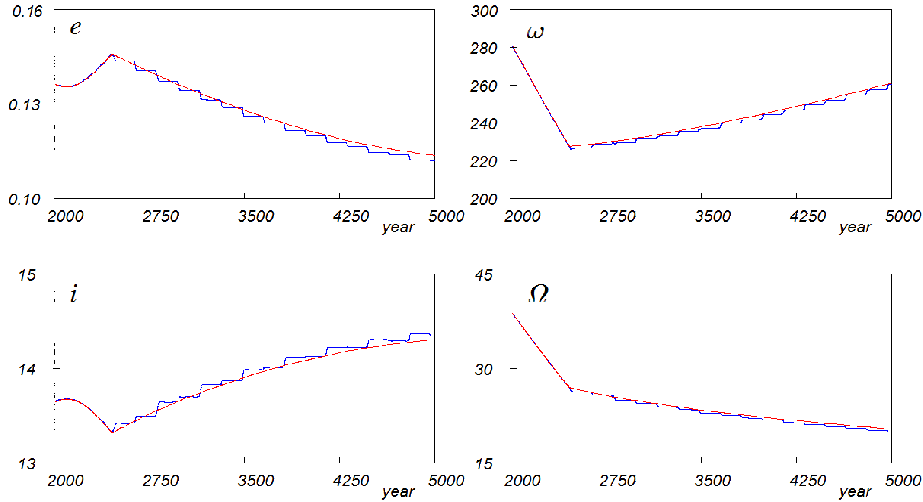}
\vglue-2.0cm
\caption{The evolution of the orbital elements of 2004GU9. Blue lines correspond
to the result of integration of the equation of motion; red lines characterise the
secular behaviour of the orbital elements according to the evolutionary equations
(\ref{eveq})}\label{aoe}
\end{figure}

\begin{figure}[htb]
\vglue3.8cm \hglue2.cm
\includegraphics[width=0.6\textwidth,keepaspectratio]{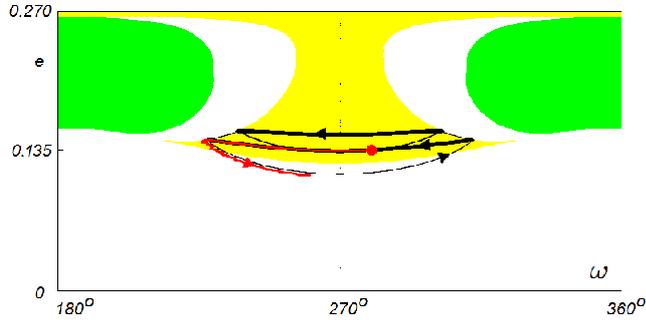}
\vglue-1.0cm
\caption{Phase portrait of the system (\ref{eveq}) at $\xi=2.87506, \Ph=0.96269$
($\sigma=0.27061$). The red line characterises the evolution of $\omega$ and $e$ in
the solution of the equations of motion with the initial
conditions corresponding to the current values of the orbital elements of 2004GU9 {\bf (marked with the
red dot)}}
\label{app}
\end{figure}

 The graphs in Fig.~\ref{aoe} allow us to compare the long term evolution of the osculating
 elements, according to Eqs.~(\ref{eveq}), with the results of a direct numerical integration. In
 Fig.~\ref{app} we present the projection of the asteroid phase trajectory onto the plane
 $\omega,\,e$ (red line). Black lines correspond to the appropriate phase trajectories of the
 system~(\ref{eveq}) when the matching at the uncertainty curve is accomplished
 according to the procedure described in
 Sec.2.4. Fig.~\ref{aoe} and Fig.~\ref{app} demonstrate that our approach based on double
 averaging of the equations of motion provides a really accurate
 description of the secular evolution.

\section*{Conclusion}

 In this paper we have considered the three body system ``Sun-planet-asteroid" in the 1:1
 mean-motion resonance. A special attention was given to the motion of the asteroid in
 a quasi-satellite orbit, when the asteroid is located (permanently or for a long enough time)
 in the vicinity of the planet, though being out of the planet's Hill sphere.

 As usual, in the mean-motion resonance three dynamical time scales can be distinguished.
 The ``fast" process corresponds to the planet and asteroid orbital motions. The
 ``semi-fast" process is a variation of the resonant phase (which, in a certain sense,
 describes the relative position of the planet and asteroid in their orbital motion).
 Finally, the ``slow" process is the secular evolution of the orbit's shape
 (the eccentricity) and orientation (the longitude of the ascending node, the inclination,
 and the argument of the pericenter).

 To study the ``slow" process, we constructed the evolutionary equations by means of numerical
 averaging over the ``fast" and ``semi-fast" motion. As a specific feature of these
 evolutionary equations, we should point out the ambiguity of their right-hand sides for some
 values of the ``slow" variables. The ambiguity appears because the averaging can be performed over the
 ``semi-fast" processes with qualitatively different properties (in other
 words, it can be done over a QS-orbit, or an HS-orbit, with the same values of the Hamiltonian $\Xi$).
 Consideration of this ambiguity
 provided us with an opportunity to predict whether the motions in QS- or HS-orbits are
 permanent or not. For non-permanent motions in QS-orbits, the conditions of capture into this
 regime and escape from it have been established.

 For clarity, our analysis was restricted to the case of the asteroid motion in orbits with
 small and intermediate values of the inclinations and eccentricities, when maximum two
 regimes of the ``semi-fast" evolution are possible. The general case can be studied similarly.

 The evolutionary equations were used to draw the phase portraits demonstrating in a clear way
 the secular effects in the asteroid motion (for example, the libration or oscillation of the
 pericenter).

 Some kind of scaling has been found in the asteroid dynamics at small inclinations and
 eccentricities. It has turned out that in this case many properties of the motion (for
 example, the duration of the QS-regime) can be established using the Hill approximation in the restricted circular
 three-body problem.
 The excessive symmetry of this approximation creates a difficulty for correct description of
 all possible modes of motion, but hopefully it can be overcomed after taking into account
 the high order terms of perturbation theory.

 To illustrate the typical rates of the orbital elements's secular evolution, the dynamics
 of the near-Earth asteroid 2004GU9 was considered. This asteroid will keep its motion in a
 QS-orbit for the next several hundred years. For better prediction, our model should be
 improved by taking into account the influence of the other planets (Venus, Mars, Jupiter,
 etc), along with the appropriate modification of the averaging procedure.

\section*{Acknowledgments}

The work was supported in part by the Russian Foundation
for Basic Research (projects 13-01-00251 and NSh-2519.2012.1) and by the Presidium of the Russian
Academy of Sciences under the scope of
the Program 22 ``Fundamental problems of Solar system
investigations". We are grateful to M.A.Vashkovyak and M.Efroimsky for reading the manuscript and
useful discussions. We thank also anonymous referees for all their corrections and suggestions.

\newpage

\section*{Appendix A. Calculation of the function $W$ and its derivatives}

Below we describe some technical details of the computational
procedure which has been applied to obtain numerically the values of
the averaged disturbing function
$$
W(\vphi,x,y,P_h)=\frac{1}{2\pi}\int_0^{2\pi}R(\vphi,x,y,P_h,\hd)\,d\hd
\eqno(A.1)
$$
and of its derivatives
$$
\frac{\partial W}{\partial \vphi},\quad \frac{\partial
W}{\partial x},\quad \frac{\partial W}{\partial y},
\quad \frac{\partial W}{\partial \Ph}
$$
with no limitations on the asteroid's eccentricity or inclination. \vskip0.1in

\noindent{\it 1.Change of the variable, over which the integration
takes place.} To start with, we replace in (A.1) the integration
over the auxiliary variable $\hd$ with integration over the
asteroid's eccentric anomaly $E$. This allows us to avoid the
necessity to solve the Kepler equation in our calculation of the
asteroid's position. The relation between the variables $\hd$ and
$E$ reads as
$$
\hd=\vphi - \omega - (E-e\sin E).
\eqno(A.2)
$$
As a result, the expressions for $W$ and its derivative can be written as
$$
W(\vphi,x,y,P_h)=-\frac{1}{2\pi}\int_0^{2\pi}R\frac{\partial
\hd}{\partial E}\,dE \eqno(A.3)
$$
$$
\frac{\partial W}{\partial
\vphi}=-\frac{1}{2\pi}\int_0^{2\pi}\frac{\partial R}{\partial
\vphi}\frac{\partial \hd}{\partial E}\,dE,\quad
\frac{\partial W}{\partial
\Ph}=-\frac{1}{2\pi}\int_0^{2\pi}\frac{\partial R}{\partial
\Ph}\frac{\partial \hd}{\partial E}\,dE,
$$
$$
\frac{\partial W}{\partial
x}=-\frac{1}{2\pi}\int_0^{2\pi}\left(\frac{\partial R}{\partial
x}\frac{\partial \hd}{\partial E} + R\frac{\partial^2
\hd}{\partial x \partial E}\right)\,dE,
$$
$$
\frac{\partial W}{\partial
y}=-\frac{1}{2\pi}\int_0^{2\pi}\left(\frac{\partial R}{\partial
y}\frac{\partial \hd}{\partial E} + R\frac{\partial^2
\hd}{\partial y \partial E}\right)\,dE,
$$
where
$$
\frac{\partial \hd}{\partial E}=-(1 - e\sin E),
$$
$$
\frac{\partial^2 \hd}{\partial E \partial x}= \frac{\partial^2
\hd}{\partial E \partial e}\cdot \frac{\partial e}{\partial x}=
\cos E \cdot \frac{\partial e}{\partial x},
$$
$$
\frac{\partial^2 \hd}{\partial E \partial y}= \frac{\partial^2
\hd}{\partial E \partial e}\cdot \frac{\partial e}{\partial y}=
\cos E \cdot \frac{\partial e}{\partial y}.
$$
The ``minus" sign before the integrals in formulae (A.3) appears due to
our intention to have the upper limit larger than the lower one.
\vskip0.1in

\noindent{\it 2. Derivatives of $R$ with respect to $\vphi,x,y,\Ph$.}
At this point, we need to introduce the uniformly rotating
heliocentric reference frame $O\xi\eta\zeta$ with the axis $O\xi$
being directed from the Sun to the planet and the axis $O\zeta$ being
aligned with the normal to the plane of the primary's orbital
motion. In this reference frame the position vector of the planet is
$\rb'=\eb_\xi$, where $\eb_\xi=(1,0,0)^T$ is the unit vector corresponding
to axis $O\xi$. Respectively the expression for disturbing
function (\ref{pf}) is reduced to
$$
R=\frac{1}{|\rb - \eb_\xi|}-(\rb,\eb_\xi).
$$
Then we get:
$$
\frac{\partial R}{\partial \vphi}=\frac{\partial R}{\partial
\rb}\frac{\partial \rb}{\partial \vphi},\quad \frac{\partial
R}{\partial x}=\frac{\partial R}{\partial \rb}\frac{\partial
\rb}{\partial x},\quad \frac{\partial R}{\partial
y}=\frac{\partial R}{\partial \rb}\frac{\partial \rb}{\partial y},
\quad \frac{\partial R}{\partial
\Ph}=\frac{\partial R}{\partial \rb}\frac{\partial \rb}{\partial \Ph},
$$
where
$$
\frac{\partial R}{\partial \rb}= - \frac{\rb - \eb_\xi}{|\rb -
\eb_\xi|^3}- \eb_\xi.
$$
\vskip0.1in

\noindent{\it 3. Position vector of the asteroid}.
Let the unit vector $\eb^*_\xi$ be directed to the pericenter of
the osculating orbit, while one more unit vector $\eb^*_\eta$ is
parallel to this orbit's minor semi-axis (and directed in the same
way as asteroid's velocity vector at the pericenter). It is not
difficult to write down the expressions for introduced unit
vectors through their projections on the axes of the reference
frame $O\xi\eta\zeta$:
$$
\eb^*_\xi=\left(
\begin{array}{c}
\cos\hd\cos\omega-\cos i \sin\hd\sin\omega \\
\sin\hd\cos\omega+\cos i \cos\hd\sin\omega \\
\sin i\sin \omega
\end{array}\right) \eqno(A.4)
$$
and
$$
\eb^*_\eta=\left(
\begin{array}{c}
-\cos\hd\sin\omega-\cos i \sin\hd\cos\omega \\
-\sin\hd\sin\omega+\cos i \cos\hd\cos\omega \\
\sin i\cos\omega
\end{array}\right) \eqno(A.5)
$$
After that the asteroid's position vector $\rb$ can be written as
the linear combination
$$
\rb(i,e,\omega,\hd,E) = \eb^*_\xi(i,\omega,\hd)
\xi^*(e,E)+\eb^*_\eta(i,\omega,\hd) \eta^*(e,E) \eqno(A.6)
$$
with the coefficients (here the orbit with $a=1$ is considered)
$$
\xi^*(e,E)=\cos E - e,\qquad \eta^*(e,E)=\sqrt{1-e^2}\sin
E.
$$

\noindent {\it 4. Derivatives of the position vector $\rb$ with
respect to $\vphi,x,y,\Ph$}. Taking into account (A.2) and (A.6) we
obtain the following relations:
$$
\frac{\partial \rb}{\partial x}=\left(\frac{\partial
\rb}{\partial \hd}\frac{\partial \hd}{\partial e}+\frac{\partial
\rb}{\partial e}\right)\frac{\partial e}{\partial x}+
\left(\frac{\partial \rb}{\partial \hd}\frac{\partial
\hd}{\partial \omega}+\frac{\partial \rb}{\partial
\omega}\right)\frac{\partial \omega}{\partial x}+ \frac{\partial
\rb}{\partial i}\frac{\partial i}{\partial x},
$$
$$
\frac{\partial \rb}{\partial y}=\left(\frac{\partial
\rb}{\partial \hd}\frac{\partial \hd}{\partial e}+\frac{\partial
\rb}{\partial e}\right)\frac{\partial e}{\partial y}+
\left(\frac{\partial \rb}{\partial \hd}\frac{\partial
\hd}{\partial \omega}+\frac{\partial \rb}{\partial
\omega}\right)\frac{\partial \omega}{\partial y}+ \frac{\partial
\rb}{\partial i}\frac{\partial i}{\partial y},
$$
$$
\frac{\partial \rb}{\partial \vphi}=\frac{\partial \rb}{\partial
\hd}\frac{\partial \hd}{\partial \vphi}=\frac{\partial
\rb}{\partial \hd},\quad
\frac{\partial \rb}{\partial \Ph}=\frac{\partial \rb}{\partial
i}\frac{\partial i}{\partial \Ph},
$$
where evidently
$$
\frac{\partial \hd}{\partial \omega}=-1,\qquad
\frac{\partial \hd}{\partial e}=\sin E.
$$

\noindent{\it 5. Derivatives of the position vector $\rb$ with
respect to $\hd$ and osculating variables $i,e,\omega$.} Direct
differentiation of (A.6) results in
$$
\frac{\partial \rb}{\partial \hd}=\frac{\partial
\eb^*_\xi}{\partial \hd}\xi^* +\frac{\partial \eb^*_\eta}{\partial
\hd}\eta^*,\qquad \frac{\partial \rb}{\partial i}=\frac{\partial
\eb^*_\xi}{\partial i}\xi^* +\frac{\partial \eb^*_\eta}{\partial
i}\eta^*, \eqno(A.7)
$$
$$
\frac{\partial \rb}{\partial \omega}=\frac{\partial
\eb^*_\xi}{\partial \omega}\xi^* +\frac{\partial
\eb^*_\eta}{\partial \omega}\eta^*, \qquad
\frac{\partial
\rb}{\partial e}=\eb^*_\xi \frac{\partial \xi^*}{\partial
e}+\eb^*_\eta\frac{\partial \eta^*}{\partial e},
$$
where
$$
\frac{\partial \xi^*}{\partial e}=1,\qquad \frac{\partial
\eta^*}{\partial e}=-\frac{e \sin E}{\sqrt{1-e^2}}.
$$
The derivatives of the unit vectors $\eb^*_\xi$ and $\eb^*_\eta$
in (A.7) can be easily evaluated from (A.4) and (A.5)
respectively. \vskip0.1in

\noindent{\it 6. Derivatives of the osculating variables
$e,i,\omega$ with respect to $x,y,\Ph$}. First we rewrite the
expressions for $e,\cos i,\sin i$ in a more compact way (compare
with (\ref{oe})):
$$
e=\frac{\sqrt{s(x^2+y^2)}}{2}, \eqno(A.8)
$$
$$
\cos i = \frac{2P_h}{s-2},\qquad \sin i =
\frac{\sqrt{(s-2)^2 - 4P_h^2}}{s-2}.
$$
Here
$$
s=4 - (x^2 + y^2).
$$
After the differentiation of the expressions (A.8) with respect
to $x,y$ we obtain:
$$
\frac{\partial e}{\partial
x}=\frac{x}{2}\left(\sqrt{\frac{s}{x^2+y^2}}-\sqrt{\frac{x^2+y^2}{s}}\right),
$$
$$
\frac{\partial e}{\partial
y}=\frac{y}{2}\left(\sqrt{\frac{s}{x^2+y^2}}-\sqrt{\frac{x^2+y^2}{s}}\right),
$$
and
$$
\frac{\partial i}{\partial x}=-\frac{4P_h
x}{(s-2)\sqrt{(s-2)^2 - 4P_h^2}},
$$
$$
\frac{\partial i}{\partial y}=-\frac{4P_h
y}{(s-2)\sqrt{(s-2)^2 - 4P_h^2}},
$$
$$
\frac{\partial i}{\partial \Ph}=-\frac{2}{\sqrt{(s-2)^2 - 4P_h^2}}.
$$
To finalise we need to find $\partial \omega/\partial x$ and
$\partial \omega/\partial y$. Taking into account in what way the
variables $x,y$ were introduced (i.e., formulae (\ref{xyd})),
the obvious relations can be written:
$$
\cos \omega = \frac{x}{\sqrt{x^2 + y^2}},\qquad \sin \omega =
-\frac{y}{\sqrt{x^2 + y^2}}. \eqno(A.9)
$$
Differentiating (A.9) we get:
$$
\frac{\partial \omega}{\partial x}= \frac{y}{x^2+y^2},\qquad
\frac{\partial \omega}{\partial y}=-\frac{x}{x^2+y^2}.
$$

\section*{Appendix B. Calculation of the leading term $\Wd$ of the averaged
disturbing function}

As it was mentioned in Sec. 4, the leading term $\Wd$ in the expression for
disturbing function (\ref{WS}) and its derivatives
$$
\frac{\partial \Wd}{\partial \xd},\quad
\frac{\partial \Wd}{\partial \yd},\quad
\frac{\partial \Wd}{\partial \vphid}
$$
can be written in terms of complete elliptic integrals $K(k)$ and $E(k)$ of the
first and second kind.

{\it B1. Calculation of $\Wd$.} We begin with a slightly modified version of the formula used
above to define $\Wd$:
$$
\Wd(\vphid,\xd,\yd)=\frac{1}{2\pi}\int^{\pi}_{-\pi}\frac{du}{\Delta(u,\vphid,\xd,\yd)},
\eqno(B.1)
$$
where $u=-\hd=\lambda'-h$,
$$
\Delta^2(u,\vphid,\xd,\yd)=\xd^2+\yd^2+\vphid^2 + 4\vphid(\xd\sin u + \yd\cos u)+
$$
$$
3(\xd\sin u + \yd\cos u)^2 + [1-(\xd^2+\yd^2)]\sin^2u.
$$
By means of the standard change of variables $\xi = \tg \frac{u}{2}$, we obtain:
$$
\Wd(\vphid,\xd,\yd)=\frac{1}{\pi}\int^{+\infty}_{-\infty}\frac{d\xi}{\sqrt{P_4(\xi)}}.
\eqno(B.2)
$$
Here $P_4(\xi)$ denotes the forth degree polynomial
$$
P_4(\xi)=d_4\xi^4 + d_3\xi^3 + d_2\xi^2 + d_1\xi + d_0
$$
with the coefficients rendered by the formulae
$$
d_0=(2\yd+\vphid)^2 + \xd^2,\quad
d_1=4\xd(2\vphid + 3\yd),\quad
d_2=2(2+5\xd^2-4\yd^2+\vphid^2),
$$
$$
d_3=4\xd(2\vphid - 3\yd),\quad
d_5=(2\yd-\vphid)^2 + \xd^2.
$$

The integral (B.1) has a finite value only in the case
$\Delta(u,\vphid,\xd,\yd)\ne 0$ for $\forall u\in [-\pi,\pi]$. It takes place
when all the roots
$\xi_1,\ldots,\xi_4$ of the equation $P_4(\xi)=0$ are not real:
$$
\xi_1=a_1+ib_1,\quad
\xi_2=a_2+ib_2,\quad\xi_3=\overline{\xi}_1,\quad\xi_4=\overline{\xi}_2\quad
(b_1>0,b_2>0).
$$
If $a_{1,2}$ and $b_{1,2}$ are known, the value of $\Wd$ is provided by the formula
$$
\Wd(\vphid,\xd,\yd)=\frac{4}{\pi\sqrt{d_4\kappa_0}}
K\left(\sqrt{\frac{\kappa_1}{\kappa_0}}\right).
$$
where
$$
\kappa_0 = (a_1-a_2)^2+(b_1+b_2)^2,\qquad \kappa_1 = (a_1-a_2)^2+(b_1-b_2)^2.
$$

{\it B2. Calculation of the derivatives of $\Wd$}. By differentiation
of (B.2) we arrive at the evident formulae
$$
\frac{\partial \Wd}{\partial \vphid}=
\sum_{k=0}^{4}\frac{\partial d_k}{\partial \vphid}X_k,\quad
\frac{\partial \Wd}{\partial \xd}=
\sum_{k=0}^{4}\frac{\partial d_k}{\partial \xd}X_k,\quad
\frac{\partial \Wd}{\partial \yd}=
\sum_{k=0}^{4}\frac{\partial d_k}{\partial \yd}X_k,
$$
where
$$
X_k=\int^{+\infty}_{-\infty}\frac{\xi^k d\xi}{P_4^{3/2}(\xi)}.
\eqno(B.3)
$$
Then we apply the change of the variable in (B.3) as
$$
\xi = a_1 - b_1\ctg\frac{\sigma - \sigma_*}{2}
\eqno(B.4)
$$
The auxiliary quantity $\sigma_*$ in (B.4) is defined by the
relations
$$
\sin \sigma_* = \frac{2(a_2-a_1)b_1}
{\sqrt{[(a_1-a_2)^2+(b_1-b_2)^2][(a_1-a_2)^2+(b_1+b_2)^2]}},
$$
$$
\cos \sigma_* = \frac{(a_1-a_2)^2+(b_1^2-b_2^2}
{\sqrt{[(a_1-a_2)^2+(b_1-b_2)^2][(a_1-a_2)^2+(b_1+b_2)^2]}},
$$
The change of variable (B.4) allows us to express $X_0,\ldots,X_4$
as linear combinations of the integrals
$$
I_k=\int_0^{\pi}\frac{\cos^k \sigma d\sigma}
{(\alpha-\beta \cos\sigma)^{3/2}}\quad (k=0,1,2)
$$
with
$$
\alpha=1+\frac{\kappa_1}{\kappa_0},\qquad
\beta = 2\sqrt{\frac{\kappa_1}{\kappa_0}}.
$$
With the aid of the standard handbook~\citet{GradshteynandRyzhik2007}, one can easily find that
$$
I_0=\frac{2}{(\alpha-\beta)\sqrt{\alpha+\beta}}
E\left(\sqrt{\frac{2\beta}{\alpha+\beta}}\right),
$$
$$
I_1=\frac{\alpha}{\beta}I_0-\frac{2}{\beta\sqrt{\alpha+\beta}}
K\left(\sqrt{\frac{2\beta}{\alpha+\beta}}\right),
$$
$$
I_2=\frac{2\alpha}{\beta}I_1-\frac{\alpha^2}{\beta^2}I_0+
\frac{2\sqrt{\alpha+\beta}}{\beta^2}
E\left(\sqrt{\frac{2\beta}{\alpha+\beta}}\right).
$$

So the relations connecting the integrals $I_k$ and $X_k$ allow
to express $X_k$ in terms of complete elliptic integrals too. Omitting the
cumbersome calculations we write down these relations in the recursive form
$$
X_0=\frac{2}{b_1^2(d_4\kappa_0)^{3/2}}\left[(1+\sin^2 \sigma_*)I_0 -
2\cos\sigma_*\cdot I_1 + \cos 2\sigma_*\cdot I_2\right],
$$
$$
X_1=a_1 X_0 +
\frac{2}{b_1(d_4\kappa_0)^{3/2}}\left[\frac{1}{2}\sin 2\sigma_*\cdot I_0 +
\sin\sigma_*\cdot I_1 - \sin 2\sigma_*\cdot I_2\right],
$$
$$
X_2=-a_1^2 X_0 + 2a_1 X_1 +
\frac{2}{(d_4\kappa_0)^{3/2}}\left[\cos^2\sigma_*\cdot I_0 -
\cos 2\sigma_*\cdot I_2\right],
$$
$$
X_3 = a_1^3 X_0 - 3a_1^2 X_1 + 3a_1 X_2
$$
$$
+\frac{2b_1}{(d_4\kappa_0)^{3/2}}\left[-\frac{1}{2}\sin 2\sigma_*\cdot I_0 +
\sin\sigma_*\cdot I_1 + \sin 2\sigma_*\cdot I_2\right],
$$
$$
X_4 =-a_1^4 X_0 + 4a_1^3 X_1 - 6a_1^2 X_2 + 4a_1 X_3 +
$$
$$
\frac{2}{b_1^2(d_4\kappa_0)^{3/2}}\left[(1+\sin^2 \sigma_*)I_0 +
2\cos\sigma_*\cdot I_1 + \cos 2\sigma_*\cdot I_2\right]
$$

{\it B3. Computation of the second derivative with respect to $\vphid$.}
To make a conclusion regarding the possibility of a QS-regime at given values of
$\xd,\yd$, we need to know the sign of $\partial^2 \Wd/\partial \vphid^2$ at $\vphid=0$
(Christou, 2000).
The answer can be obtained through the use of the formula
$$
\left.\frac{\partial^2 \Wd}{\partial \vphid^2}\right|_{\vphid=0}=
-\frac{1}{\pi}I_0 + \frac{6}{\pi}(x^2+y^2)J_0
$$
$$
+\frac{6}{\pi}\frac{\left[(y^2-x^2)(1+2x^2-4y^2)-12x^2y^2\right]J_1}{\sqrt{20x^2y^2+(1+2x^2)^2+(1-4y^2)^2-1}}
$$
where
$$
J_0=\int_0^{\pi}\frac{d\xi}{(\alpha - \beta\cos\xi)^{5/2}} =
\frac{2\sqrt{\alpha+\beta}}{3(\alpha^2-\beta^2)^2}
\left[4\alpha E\left(\sqrt{\frac{2\beta}{\alpha+\beta}}\right)-(\alpha-\beta)
K\left(\sqrt{\frac{2\beta}{\alpha+\beta}}\right)\right],
$$
$$
J_1=\int_0^{\pi}\frac{cos\xi d\xi}{(\alpha - \beta\cos\xi)^{5/2}} =
\frac{\alpha}{\beta}J_0 - \frac{1}{\beta}I_0,
$$
$$
\alpha=\frac{1}{2}+2(\xd^2+\yd^2),\qquad
\beta=\frac{1}{2}\sqrt{20x^2y^2+(1+2x^2)^2+(1-4y^2)^2-1}.
$$

{\bf We can add also that the value of $\Wd$ at $\vphid=0$ is provided by the formula}
$$
\Wd = \frac{1}{\pi}\frac{2}{\sqrt{\alpha + \beta}}
K\left(\sqrt{\frac{2\beta}{\alpha+\beta}}\right).
\eqno(B.5)
$$
{\bf Up to notations the formula (B.5) coincides with the formula (46) for the secular potential
in Namouni (1999) which was used there to study the limiting case of QS-orbits with
non-oscillating resonant phase $\vphi$.}

\end{document}